\documentclass[a4paper,11pt]{article}

\usepackage{jheppub} 
\usepackage{graphicx,float,wrapfig,subfigure}
\usepackage{amsfonts,amsmath,amssymb,amstext}
\usepackage{latexsym}
\usepackage{bm}
\usepackage{color}
\usepackage[normalem]{ulem}
\usepackage{MnSymbol}
\usepackage[colorlinks=true,linktocpage=true,linkcolor=blue,citecolor=blue]{hyperref}
\usepackage{slashed}
\usepackage{orcidlink}    

\usepackage[T1]{fontenc} 

\newcommand{\sign}{ \mbox{sign}}
\renewcommand{\Im}{\mbox{Im}}

\newcommand{\tr}{ \mbox{tr}}
\definecolor{red}{rgb}{0.7,0,0}
\definecolor{green}{rgb}{0,0.5,0}

\title{Neutrino energy and momentum emission from magnetized dense quark matter}

\author[a]{Ritesh Ghosh\orcidlink{0000-0002-6740-7038}}
\author[a,b]{and Igor A. Shovkovy\note{Corresponding author.}\orcidlink{0000-0002-5230-6891}}

\affiliation[a]{College of Integrative Sciences and Arts, Arizona State University,\\
Mesa, Arizona 85212, U.S.A.}
\affiliation[b]{Department of Physics, Arizona State University, \\
Tempe, Arizona 85287, U.S.A.}

\emailAdd{Ritesh.Ghosh@asu.edu}
\emailAdd{igor.shovkovy@asu.edu}

\abstract{Using first-principles field-theoretic methods, we investigate neutrino emission from strongly magnetized dense quark matter under conditions relevant to compact stars. We develop a customized approximation that fully accounts for the Landau-level quantization of electron states while neglecting such quantization for quarks. This approach is well-justified in dense quark matter, where the chemical potentials of up and down quarks significantly exceed those of electrons. Our analysis provides a detailed exploration of the influence of strong magnetic fields on neutrino emission, including both the modification of the total emission rate and the emergence of emission asymmetry relative to the magnetic field direction. We further examine the role of temperature in smoothing the oscillatory behavior of neutrino emission as a function of magnetic field strength. Additionally, we study the interplay between the Landau-level quantization of electrons and the Fermi-liquid effects of quarks in modifying the phase space of relevant weak processes. Finally, we briefly discuss the broader implications of magnetic fields on stellar cooling processes and the potential contribution of asymmetric neutrino emission to pulsar kicks.}

\arxivnumber{2501.03318}

\date{March 17, 2025}

\begin{document} 
\maketitle
\flushbottom

\section{Introduction}
\label{sec:Introduction}

Compact stars may host exotic forms of matter at their cores, including quark matter, which is an extreme state of baryonic matter where quarks are no longer confined within nucleons but exist as deconfined particles. The presence of quark matter in stellar cores is of considerable interest, as it could fundamentally alter the physical properties and observable characteristics of compact stars~\cite{Baym:2017whm,Annala:2023cwx}. Additionally, many compact stars exhibit exceptionally strong magnetic fields, with surface field strengths reaching up to $10^{15}~\mbox{G}$ or higher in the case of magnetars \cite{Turolla:2015mwa,Kaspi:2017fwg}. These fields can be even stronger in the dense core regions \cite{Lai_1991}, making the study of their effects on the underlying matter essential for understanding stellar behavior.

One of the critical features of compact stars is their cooling behavior, which can be tested and constrained through observational data. This cooling is primarily governed by the interplay between the specific heat of the stellar matter and the rate of energy loss. Generally, the dominant energy loss mechanism is neutrino emission, which originates from the entire volume of the star.  In contrast, photon emission, which comes from the stellar surface only, plays a secondary role in the overall energy balance despite its importance for observations (for reviews, see refs.~\cite{Baym:2017whm,Yakovlev:2004iq}).

The influence of magnetic fields on neutrino emission processes has been a subject of active research \cite{Baiko:1998jq,Riquelme:2005ac,Potekhin:2015qsa}, as magnetic fields can modify emission rates and introduce anisotropies in the emission patterns. The former affects the thermal evolution of the star, while the latter may contribute to the pulsar kicks  \cite{Lai:1998sz,Arras:1998cv,Sagert:2007as}. Most studies, however, have concentrated on the effects of magnetic fields on nuclear matter \cite{Baiko:1998jq,Riquelme:2005ac,Potekhin:2015qsa,Potekhin:2017ufy,Dehman:2022rpa,Tambe:2024usx,Kumamoto:2024jiq}, revealing phenomena such as a reduced threshold for direct Urca processes and an oscillatory dependence of emission rates on the magnetic field strength. In contrast, investigations of strongly magnetized dense quark matter remain sparse \cite{Belyaev:2017nos}. While some studies have tried addressing momentum emission \cite{Sagert:2007as,Ayala:2018kie,Ayala:2024wgb}, they relied on oversimplified models that use the degree of electron spin polarization to estimate neutrino emission asymmetry. Thus, it is fair to say that a comprehensive analysis of the underlying physics is still lacking. 

In this work, we address this knowledge gap by using first-principles field-theoretic methods to calculate both the energy and net longitudinal momentum emission rates, providing a rigorous foundation for understanding neutrino emission in magnetized dense quark matter. Recognizing that the chemical potentials of quarks are significantly larger than that of electrons, we develop a self-consistent framework that incorporates the Landau-level quantization of electron states while neglecting such quantization for quarks. Our analysis accounts for Fermi-liquid corrections to the quark dispersion relations \cite{Baym:1975va}, which is consistent with the formalism used in the absence of a magnetic field~\cite{Iwamoto:1980eb,Iwamoto:1982zz,Schafer:2004jp}. Furthermore, we examine in detail how the neutrino emission rate and its directional asymmetry depend on the magnetic field strength and temperature. By investigating the impact of strong magnetic fields, we aim to shed light on the thermal evolution of magnetars and the origin of pulsar kicks, as well as provide guidance for identifying potential observational signatures of magnetized quark matter in their cores.

\begin{figure}[t]
\centering
  \subfigure[]{\includegraphics[width=0.375\textwidth]{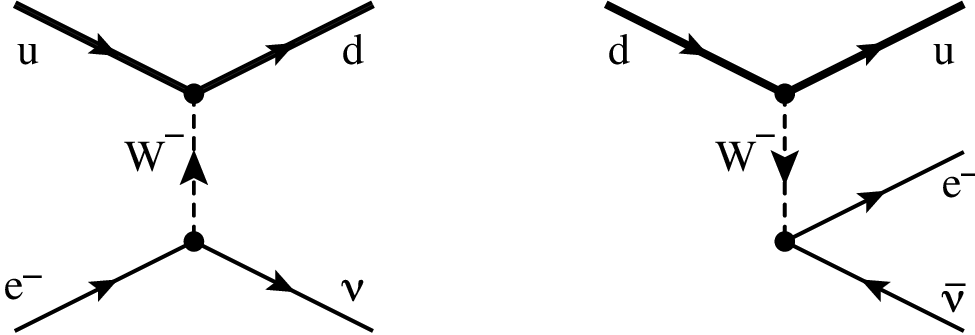}}
  \hspace{0.1\textwidth}
  \subfigure[]{\includegraphics[width=0.25\textwidth]{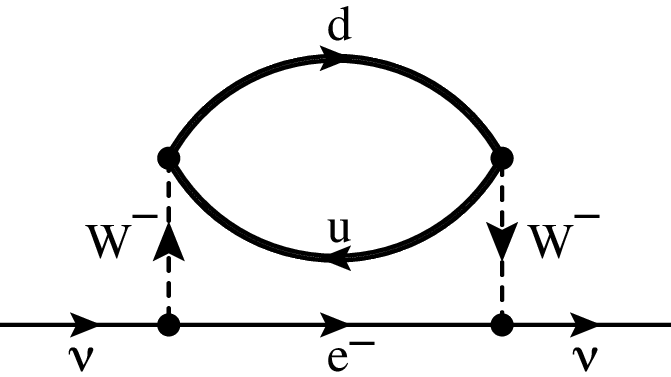}}
\caption{(a) Direct Urca processes dominating (anti-)neutrino emission from dense quark matter. 
(b) The neutrino self-energy diagram which is used to calculate the (anti-)neutrino emission rate.}
\label{fig.NuSelfEnergy}
\end{figure}

Neutrino emission from unpaired dense quark matter primarily comes from the direct Urca processes, which involve weak interactions between quarks and leptons~\cite{Iwamoto:1980eb,Iwamoto:1982zz,Schafer:2004jp}. The two underlying processes are shown in figure~\ref{fig.NuSelfEnergy}(a): electron capture by an up quark, producing a down quark and an electron neutrino ($u+ e^- \rightarrow d+\nu_e$), and the decay of a down quark into an up quark, accompanied by the emission of an electron and an electron antineutrino ($d\rightarrow u +e^- + \bar{\nu}_e$). Because neutrinos and antineutrinos have extremely long mean-free paths, they escape the star without rescattering, efficiently carrying away energy and acting as a dominant cooling mechanism for dense quark matter in a stellar core. The dependence of emission rates on the density and temperature of quark matter has been well established since Iwamoto’s pioneering works in the 1980s~\cite{Iwamoto:1980eb,Iwamoto:1982zz}. In this study, we demonstrate that strong magnetic fields can profoundly alter the phase space of electrons, leading to significant changes in neutrino emission rates and introducing anisotropies in the emission pattern.

This paper is organized as follows. In section~\ref{sec:formalism}, we describe the general framework for calculating neutrino emission from unpaired dense quark matter in a background magnetic field using field-theoretic methods. Specifically, we apply the Kadanoff-Baym formalism to derive the rate of change of the (anti-)neutrino distribution function due to electron capture and down quark decay. Recognizing that the lepton and quark contributions naturally factorize in the final expression, we implement a series of well-justified approximations to ensure efficient and reliable computation of energy and net momentum emission rates.
The primary numerical results are presented in section~\ref{sec:Numerics}, where we investigate the dependence of emission rates on magnetic field strength, temperature, and electron chemical potential. We also explore the partial contributions from electrons occupying different Landau levels near the Fermi surface, uncovering several intriguing features of energy and momentum emission. Furthermore, we demonstrate that the Landau-level quantization of electron states plays a role similar to that of the Fermi-liquid corrections in quark dispersion relations. Using our numerical data, we briefly discuss the implications of neutrino emission for cooling rate and pulsar kick velocities. Finally,  section~\ref{sec:Summary} summarizes the key findings of our study. For completeness, additional technical details, including the derivation of the benchmark neutrino emission rate in the absence of a magnetic field, are provided in several Appendices at the end of the paper.

\section{General Formalism}
\label{sec:formalism}

To calculate the (anti-)neutrino emission rate, we employ the Kadanoff-Baym transport equation for neutrinos, following the approach used in refs.~\cite{Sedrakian:1999jh,Schmitt:2005wg,Jaikumar:2005hy}. For this study, we consider unpaired dense quark matter composed of the two lightest quark flavors, up and down. We further assume that the quark matter temperature is below the threshold for neutrino trapping  \cite{Prakash:1996xs}, implying that neutrino chemical potential vanishes ($\mu_\nu =0$). To maintain electrical neutrality, a small but nonzero density of electrons must be present in the system. In $\beta$-equilibrium, the chemical potentials of the down and up quarks ($\mu_d$ and $\mu_u$) are related to the electron chemical potential ($\mu_e$) by the relation $\mu_d = \mu_u + \mu_e$.

\subsection{Kadanoff-Baym transport equation for neutrinos}

Without loss of generality, we assume that dense quark matter in the stellar core is spatially homogeneous. Then, the out-of-equilibrium neutrino Green functions $G^{\lessgtr}_\nu(t,P_\nu)$ depends only on time and can be expressed in terms of the (left-handed) neutrino distribution function as follows \cite{Sedrakian:1999jh,Schmitt:2005wg}:
{\small
\begin{eqnarray}
i G^{<}_\nu(t,P_\nu) &=& -\frac{\pi}{p_\nu}\frac{1-\gamma_5}{2} (\gamma^\lambda P_{\nu,\lambda}+\mu_\nu \gamma_0) \left\{f_\nu(t,\bm{p}_\nu)\delta(p_{\nu,0}+\mu_\nu-p_\nu) -\left[1-f_{\bar \nu}(t,-\bm{p}_\nu)\right]\delta(p_{\nu,0}+\mu_\nu+p_\nu)\right\} , \nonumber\\
&&  \label{G-less} \\
i G^{>}_\nu(t,P_\nu) &=& \frac{\pi}{p_\nu} \frac{1-\gamma_5}{2} (\gamma^\lambda P_{\nu,\lambda}+\mu_\nu \gamma_0) \left\{\left[1-f_\nu(t,\bm{p}_\nu)\right] \delta(p_{\nu,0}+\mu_\nu-p_\nu)-f_{\bar \nu}(t,-\bm{p}_\nu)\delta(p_{\nu,0}+\mu_\nu+p_\nu)\right\} .\nonumber\\
&& \label{G-greater}
\end{eqnarray}}
Here, by definition, $P_\nu =(p_{\nu,0},\bm{p}_\nu)$ is the neutrino four-momentum and $p_\nu =|\bm{p}_\nu|$ is the magnitude of its three-momentum. Since we consider quark matter without neutrino trapping, we set $\mu_\nu =0$ in this study. The Green functions satisfy the following equation:
\begin{equation}
i \partial_t \mbox{Tr}[\gamma^0 G_\nu^{<} (t,P_\nu)] = -\mbox{Tr}[G_\nu^{>} (t,P_\nu)\Sigma^{<}_\nu(t,P_\nu)-\Sigma^{>}_\nu(t,P_\nu)G_\nu^{<} (t,P_\nu)].
\label{KB-kinetic-eq}
\end{equation}
In essence, this is a kinetic equation that relates the time derivative of the neutrino distribution function, appearing inside $G_\nu^{<} (t,P_\nu)$ on the left-hand side, to an implicit collision integral expressed in terms of the neutrino self-energy functions $\Sigma^{\lessgtr}_\nu(t,P_\nu)$. Note that the time dependence on the right-hand side of Eq.~(\ref{KB-kinetic-eq}) can be neglected. This approximation is justified because the macroscopic evolution of dense quark matter occurs much more slowly than the characteristic timescale of weak interactions.

To derive the neutrino self-energy, we use the low-energy Fermi theory of weak interactions. The interaction Lagrangian is given by \cite{Weinberg:1996kr}
\begin{equation}
\mathcal{L}=\frac{G_F \cos \theta_C}{\sqrt{2}} \bar u \gamma^\mu(1-\gamma_5) d \, \bar{e} \gamma_\mu (1-\gamma_5)\nu_e,
\label{interaction-Lagrangian}
\end{equation}
where $G_F\approx 1.166\times 10^{-11}~\mbox{MeV}^{-2}$ is the Fermi coupling constant and $\theta_C$ is the Cabibbo angle (note that $\cos^2\theta_C \approx 0.948$). Within this framework, the neutrino self-energy is expressed as
\begin{equation}
\Sigma_\nu^{\lessgtr}(P_\nu) = i  \frac{G_F^2\cos^2\theta_C}{2} \int \frac{d^4 Q}{(2\pi)^4} \gamma^\delta (1-\gamma^5)\bar{S}_e^{\lessgtr}(P_\nu + Q)\gamma^\sigma(1-\gamma^5)\bar{\Pi}^{\gtrless }_{\delta\sigma}(Q) , 
\label{Sigma-gtr}
\end{equation}
and the corresponding Feynman diagram is shown in figure~\ref{fig.NuSelfEnergy}(b). 

Let us recall that, in the presence of a background magnetic field, the coordinate-space representations of Green functions for charged particles (such as electrons and W-bosons) take the following forms: $S_e^{\lessgtr}(u,u^{\prime})=e^{i\Phi(u,u^{\prime})}\bar{S}_e^{\lessgtr}(u-u^{\prime})$ and $\Pi^{\lessgtr}_{\delta\sigma}(u^{\prime},u)=e^{-i\Phi(u,u^{\prime})}\bar{\Pi}^{\lessgtr}_{\delta\sigma}(u^{\prime}-u)$, respectively, where $\Phi(u,u^{\prime})$ is the well-known Schwinger phase and $u=(t,\bm{r})$ represents the spacetime coordinates. While the Schwinger phases formally break the translational invariance of the Green functions, they cancel out in the product of the two functions. This cancellation is critical, as the resulting combination is the only one that contributes to the leading-order result for the neutrino self-energy in Eq.~(\ref{Sigma-gtr}). Moreover, the final expression closely resembles that in the case without a magnetic field.

With the Schwinger phases cancelled, the neutrino self-energy in Eq.~(\ref{Sigma-gtr}) is written as usual in terms of the Fourier transforms of the translationally invariant parts of the electron propagator and the $W$-boson self-energy, i.e.,
\begin{eqnarray}
\bar{S}_e^{\lessgtr}(P_e) &=& \int  d^4 u e^{i P_e^\lambda u_\lambda }\bar{S}_e^{\lessgtr}(u) ,
\label{S-Fourier} \\
\bar{\Pi}^{\gtrless}_{\delta\sigma}(Q)&=& \int  d^4 u e^{i Q^\lambda u_\lambda } \bar{\Pi}^{\gtrless}_{\delta\sigma}(u) .
\label{Pi-Fourier}
\end{eqnarray}
Despite its conventional form, Eq.~(\ref{Sigma-gtr}) fully incorporates the effects of the background magnetic field. Compared to the zero-field case, however, the transverse components of the ``four-momenta'' $P_e$ and $Q$, introduced via the Fourier transforms (\ref{S-Fourier}) and (\ref{Pi-Fourier}), do not correspond to the physical momenta of the electron and the $W$-boson. Indeed, for charged particles, the components of momenta perpendicular to the magnetic field are not well-defined quantum numbers. Only the time-like component and the longitudinal component along the direction of the magnetic field correspond to the physical energies and conserved momenta of particles.

The Fourier transforms of the (translationally invariant parts of) electron Green functions can be written as follows:
\begin{eqnarray}
i\bar{S}_e^{>}(P_e)&=&[1-n_F(p_{e,0})]A_{e}(p_{e,0}+\mu_{e},\mathbf{p}_e), \\
i\bar{S}_e^{<}(P_e)&=&-n_F(p_{e,0})A_{e}(p_{e,0}+\mu_{e},\mathbf{p}_e),
\end{eqnarray}
where $n_{F}(p_0) = 1/\left[\exp(p_0/T)+1\right]$ is the Fermi-Dirac distribution and $A_{e}(p_{e,0}+\mu_{e},\mathbf{p}_e)$ is the spectral function. Here our convention is such that the quasiparticle electron energy $p_{e,0}$ is measured from the Fermi surface. The spectral function in the Landau-level representation is derived in appendix~\ref{app:E-propagator}. Its explicit expression reads
\begin{eqnarray}
A_{e} (p_{e,0}+\mu_{e},\bm{p}) &=& 2\pi  e^{-p_{e,\perp}^2\ell^2}  \sum_{\lambda=\pm}\sum_{n=0}^{\infty} 
\frac{(-1)^n}{E_{e,n}}\Big\{
\left[E_{e,n}\gamma^{0} 
-\lambda  p_{e,z}\gamma^3+ \lambda m_e \right]
\left[{\cal P}_{+}L_n\left(2 p_{e,\perp}^2\ell^2\right) \right.
\nonumber\\
&-&\left. {\cal P}_{-}L_{n-1}\left(2 p_{e,\perp}^2\ell^2\right)\right] 
+2\lambda  (\bm{p}_{e,\perp}\cdot\bm{\gamma}_\perp) L_{n-1}^1\left(2 p_{e,\perp}^2\ell^2\right)
\Big\}\delta(p_{e,0} +\mu_e -\lambda E_{e,n}) ,\nonumber\\
&&
\label{spectral-density}
\end{eqnarray}
where $ E_{e,n}=\sqrt{2n|eB|+p_{e,z}^2+ m_e^2 }$ are the Landau-level energies, $\ell =1/\sqrt{|eB|}$ is the magnetic length for the electron, ${\cal P}_{\pm}=(1\pm i s_\perp \gamma^1\gamma^2)/2$ are the spin projectors, $s_\perp = \mbox{sign}(eB)$, and $L_n^{\alpha}\left(z\right)$ are the generalized Laguerre polynomials \cite{Gradshteyn:1943cpj}. 

Regarding the $W$-boson self-energies, it is convenient to express them as \cite{2000tft..book.....L}
\begin{eqnarray}
 i\Pi_{\delta\sigma}^>(Q)&=& 2[1+n_B(q_0)] \Im \left[\Pi^R_{\delta\sigma}(Q)\right],\\
 i\Pi_{\delta\sigma}^<(Q)&=&2n_B(q_0)\Im \left[\Pi^R_{\delta\sigma}(Q)\right],
\end{eqnarray}
where $n_B(q_0) =1/\left[\exp(q_0/T)-1\right]$ is the Bose-Einstein distribution function and $\Pi^R_{\delta\sigma}(Q)$ denotes the retarded Green function. As seen from figure~\ref{fig.NuSelfEnergy}(b), this self-energy is determined by the one-loop quark diagram. 

Since quarks carry electrical charges, the effects of the background magnetic field should be formally included in their propagators when calculating the retarded $W$-boson self-energy. However, we note that the quark chemical potentials $\mu_u$ and $\mu_d$ are on the order of $300~\mbox{MeV}$, which are much larger than the energy scale set by the magnetic field, namely $\sqrt{|eB|} \lesssim 25~\mbox{MeV}$, assuming realistic field strengths below $10^{17}~\mbox{G}$. In this case, a large number of quark Landau levels (on the order of hundreds)  are occupied, effectively reducing the impact of Landau-level quantization near the Fermi surface. Indeed, a simple estimate for the energy spacing $\Delta \epsilon_n$ between neighboring Landau levels at the Fermi surface scales as $ |eB|/\mu_f$ (where $f=u,d$ labels the quark flavors). This spacing is sufficiently small to be negligible compared to the effects of temperature and the Landau level widths. Therefore, the effect of the magnetic field can be safely neglected in the derivation of the $W$-boson self-energy $\Pi^R_{\delta\sigma}(Q)$. 

Unlike quarks, electrons require a more careful treatment due to a more pronounced role of Landau-level quantization. This difference stems from the considerably smaller electron chemical potential $\mu_e$, whose typical values are on the order of $50~\mbox{MeV}$. Consequently, our analysis incorporates the precise Landau-level structure of the electron states, as described by the spectral function given in Eq.~(\ref{spectral-density}).
 
\subsection{Neutrino-number production rate}

Substituting the spectral representations of the neutrino and electron Green functions, as well as the $W$-boson self-energy, into the kinetic equation (\ref{KB-kinetic-eq}), we derive the following expression for neutrino-number production rate:
\begin{eqnarray}
\frac{\partial f_\nu(t,\bm{p}_\nu)}{\partial t} &=& -\frac{G_F^2\cos^2\theta_C}{2 }
\sum_{\lambda=\pm}\sum_{n=0}^{\infty} (-1)^n
\int \frac{d^3\bm{p}_{e} e^{-p_{e,\perp}^2\ell^2} }{(2\pi)^3 p_\nu E_{e,n}}
n_F(E_{e,n}-\mu_e)  \nonumber\\
&& \times 
n_B(p_\nu+\mu_e-E_{e,n})  L_{n,\lambda}^{\delta\sigma}(\bm{p}_e,\bm{p}_\nu)  \Im \left[ \Pi^R_{\delta\sigma}(Q) \right] ,
\label{rate-01}
\end{eqnarray}
where $Q\equiv(E_{e,n} -\mu_e-p_\nu,\bm{p}_e-\bm{p}_\nu)$. The distribution function for antineutrinos 
satisfies a similar equation. The Landau-level dependent lepton tensor is defined by 
\begin{eqnarray}
L^{\delta\sigma}_{n,\lambda}(\bm{p}_e,\bm{p}_\nu)&=&\mbox{Tr}\Big[\Big\{
\left(E_{e,n}\gamma^{0} -\lambda  p_{e,z}\gamma^3+ \lambda m_e \right) 
\left[{\cal P}_{+}L_n\left(2 p_{e,\perp}^2\ell^2\right)
-{\cal P}_{-}L_{n-1}\left(2 p_{e,\perp}^2\ell^2\right)\right] \nonumber\\
&&
+2\lambda  (\bm{p}_{e,\perp}\cdot\bm{\gamma}_\perp) L_{n-1}^1\left(2 p_{e,\perp}^2\ell^2\right)
\Big\}\gamma^\sigma(1-\gamma^5)(\gamma_0 p_\nu-\bm{\gamma}\cdot \bm{p}_\nu)\gamma^\delta (1-\gamma^5)\Big].
\label{Lmunu}
\end{eqnarray}
It is worth noting that a result similar to Eq.~(\ref{rate-01}) can also be obtained in the absence of a magnetic field  \cite{Schmitt:2005wg}. In that case, the contraction of the lepton and quark tensors, $L_{n,\lambda}^{\delta\sigma}(\bm{p}_e,\bm{p}_\nu)  \Im \left[ \Pi^R_{\delta\sigma}(Q) \right]$, appearing on the right-hand side of the equation, includes the squared scattering amplitude for the relevant weak process, which depends on the particle momenta as follows: $|{\cal M}|^2 \propto(\bar{P}\cdot P_\nu)(\bar{K}\cdot P_e)$ \cite{Iwamoto:1980eb,Iwamoto:1982zz}, where $\bar{P}$ and $\bar{K}$ are the four-momenta of the down and up quarks, respectively. In the presence of a background magnetic field, the corresponding contraction is derived in appendix~\ref{sec-L-Pi}. As seen from Eq.~\eqref{L-Im-Pi}, the final expression is significantly more complicated. However, it also contains a similar combination $\left(\bar{P}\cdot P_\nu\right) \left(\bar{K}\cdot Y_e\right)$, where the electron four-momentum is replaced by $Y_e$, which incorporates the complete information about the electron’s Landau-level wave function. The explicit expression for the components of $Y_e$ are given in Eqs.~(\ref{Y0-app}) through (\ref{Yxy-app}). 

By making use of the explicit expression for $L_{n,\lambda}^{\delta\sigma}(\bm{p}_e,\bm{p}_\nu)  \Im \left[ \Pi^R_{\delta\sigma}(Q) \right]$ given in Eq.~\eqref{L-Im-Pi}, we derive
\begin{eqnarray}
\frac{\partial f_\nu(t,\bm{p}_\nu)}{\partial t} &=& \frac{N_c  G_F^2\cos^2\theta_C}{ 2  \pi^4}
\sum_{n=0}^{\infty} (-1)^n
\int \frac{pkdk d^3\bm{p}_{e} e^{-p_{e,\perp}^2\ell^2} }{v_Fq p_\nu E_{e,n}E_k E_p} \Theta(p_e) 
n_F(E_{e,n}-\mu_e) n_F( \mu_d-E_p)\nonumber\\
&&\times  n_F(E_k-\mu_u)  \Bigg\{ \left[ E_p  p_{\nu,0}- \left(1+ \frac{k}{q} \cos\theta_{eu}\right) (\bm{q}\cdot \bm{p}_\nu) \right]\left( E_k Y_{e,0}  -\frac{k}{q}  \cos\theta_{eu} (\bm{q}\cdot \bm{Y}_{e}) \right) 
\nonumber\\
&&
+\frac{1-\cos^2\theta_{eu}}{2}\left( k^2 (\bm{p}_\nu\cdot \bm{Y}_{e}) -\frac{k^2}{q^2}(\bm{p}_\nu\cdot \bm{q})(\bm{Y}_{e}\cdot \bm{q})\right) 
\Bigg\}  ,
\label{rate-02}
\end{eqnarray}
where we took into account the energy conservation relation $p_\nu+\mu_e-E_{e,n}= E_k-\mu_u-E_p+ \mu_d$, as well as the following identity for the distribution functions:
\begin{equation}
n_B(E_k-\mu_u-E_p+\mu_d)  \left[n_F( E_p- \mu_d)-n_F(E_k-\mu_u)\right] =n_F( \mu_d-E_p)n_F(E_k-\mu_u).
\end{equation}
In deriving Eq.~(\ref{rate-02}), we assumed that the emission is dominated by neutrinos with energies comparable to the temperature and the underlying weak processes involve primarily the quark states in a close vicinity of their Fermi surfaces. As in the zero-field case \cite{Iwamoto:1980eb,Iwamoto:1982zz}, these well-justified assumptions lead to important kinematic constraints: the quark momenta $\bm{k}$ and $\bm{p}$, as well as the electron pseudo-momentum $\bm{p}_e$, must be approximately parallel to each other. 

If taken at face value, the collinearity constraint significantly limits the neutrino emission rate, especially at low temperatures. However, as Iwamoto demonstrated in the 1980s \cite{Iwamoto:1980eb,Iwamoto:1982zz}, this extreme constraint is naturally alleviated when Fermi-liquid corrections to the quark dispersion relations are taken into account. These corrections reduce the Fermi momenta and velocities of quarks, leading to the following modified energy relations: $E_{p,f}=\mu_f+v_F(p-p_F)$, where $f=u,d$ labels the quark flavors, $p_F=v_F\mu_f$ is the Fermi momentum, $v_F= 1-\kappa$ is the Fermi velocity, and $\kappa = 2\alpha_s/(3\pi)$  \cite{Baym:1975va,Schafer:2004jp}. 

\begin{figure}[t]
\centering
\subfigure[]{\includegraphics[height=0.3\textwidth]{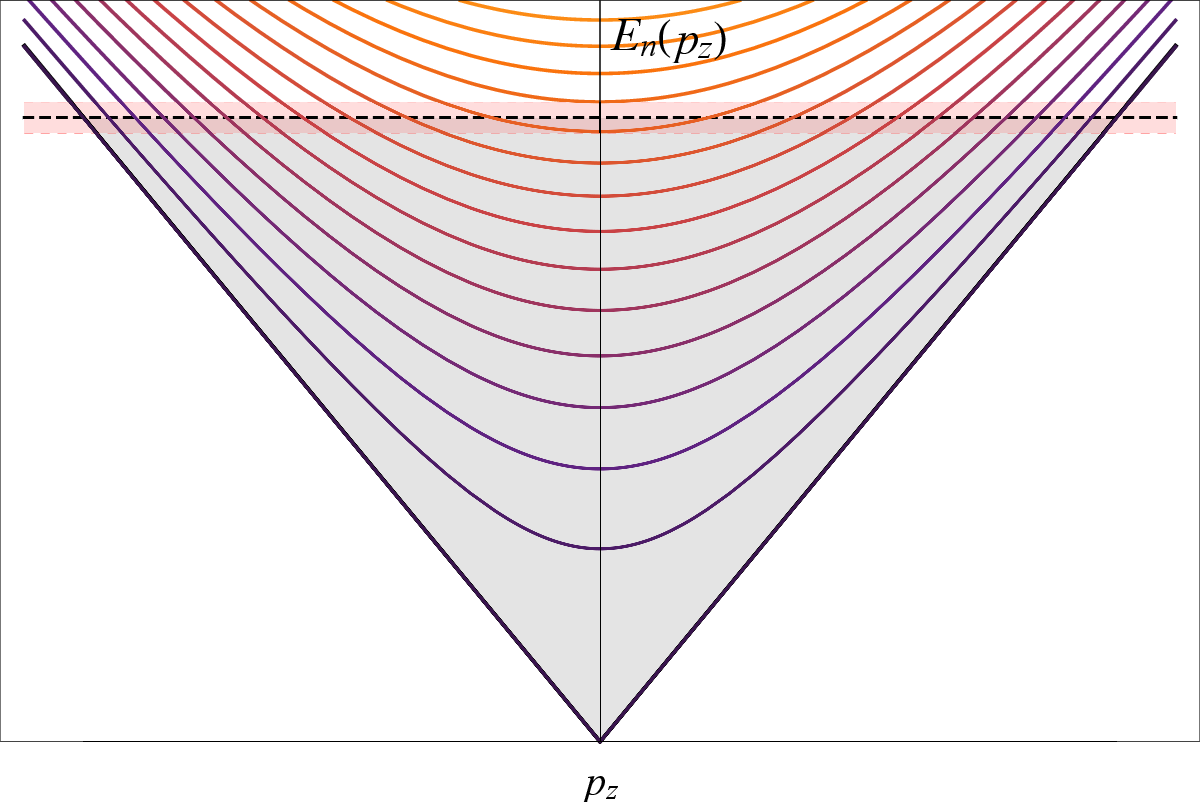}}
  \hspace{0.12\textwidth}
\subfigure[]{\includegraphics[height=0.3\textwidth]{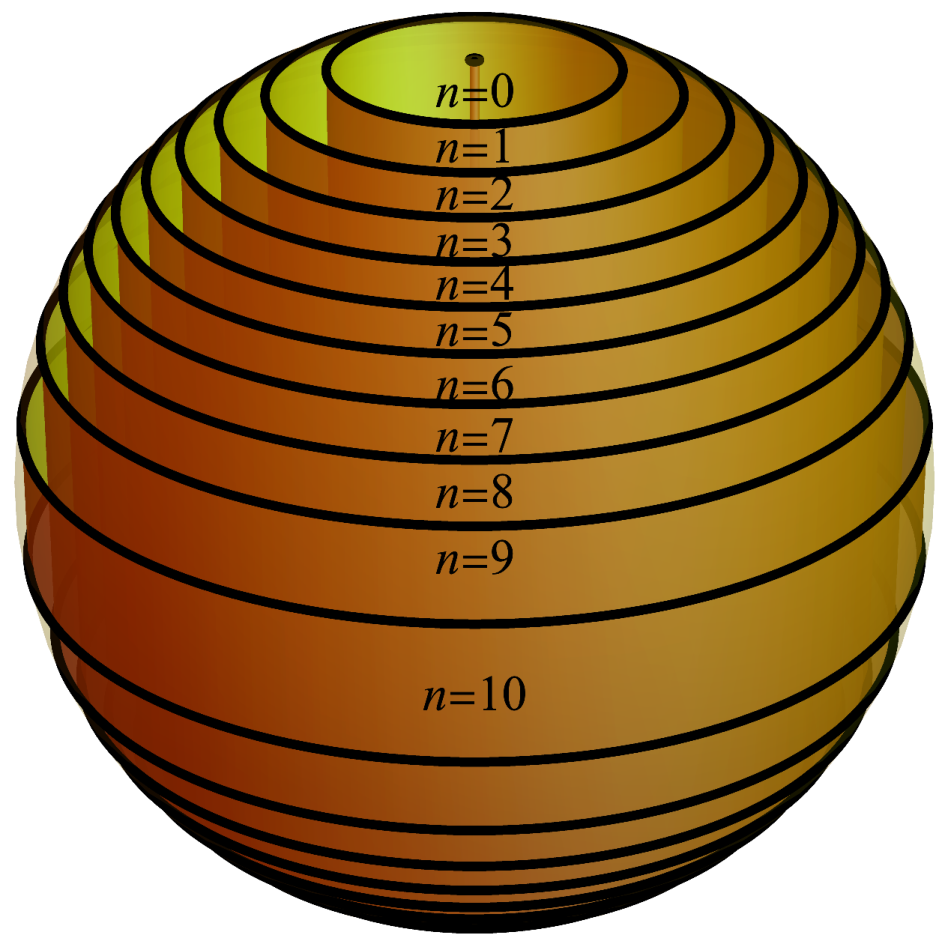}}
\caption{Schematic illustration of the electron's Landau-level spectrum (a) and the Fermi surface (b) in the presence of a strong magnetic field. (a) The Fermi level ($\mu_e$) is shown by the dashed black line and its vicinity ($\mu_e\pm T$) is highlighted with a light red band. (b) The occupied states in the Landau levels are represented as cylinders of radius $\sqrt{2n|eB|}$, extending along the vertical $p_{e,z}$ axis from $p^{(-)}_{z,F}$ to $p^{(+)}_{z,F}$.}
\label{fig:Fermi-surface}
\end{figure}

After including the Fermi-liquid corrections, one finds that the three momenta are not collinear any more. Instead, the angle between the electron pseudo-momentum  $\bm{p}_e$ and the up quark momentum $\bm{k}$ is approximately determined by 
\begin{equation}
\cos\theta_{eu}  \simeq \frac{v_F^2 (\mu_d^2-\mu_u^2)-p_e^2}{2v_F\mu_u p_e} .
\label{cos-theta}
\end{equation}
This kinematic constraint has been taken into account in deriving Eq.~(\ref{rate-02}) when performing the angular integrations associated with the direction of the up-quark momentum $\bm{k}$. To ensure that $|\cos\theta_{eu}| \leq 1$, the range of allowed values for the pseudo-momentum $p_e$ is restricted to $v_F\mu_e \leq p_e \leq v_F(\mu_d+\mu_u)$. This inequality is explicitly enforced in Eq.~(\ref{rate-02}) through the inclusion of unit-step functions inside $\Theta(p_e) $, which is defined as follows:
\begin{equation}
 \Theta(p_e)  \equiv \theta\left(p_e-v_F\mu_e\right)\theta\left[v_F(\mu_d+\mu_u)-p_e\right] .
\label{Theta-pe}
\end{equation}
While the effects of Fermi-liquid corrections resemble those in the zero-field case, overviewed briefly in appendix~\ref{app:emissionB0}, there are important differences. Specifically, one cannot assume in Eq.~(\ref{cos-theta}) that $p_e \equiv \sqrt{p_{e,\perp}^2+p_{e,z}^2}$ is approximately equal to the electron's Fermi momentum. In turn, the integration over $p_{e}$ in Eq.~(\ref{rate-02}) is not necessarily dominated by the region near $p_{e}\approx \mu_e$. This is because, in the presence of a magnetic field, $\bm{p}_{e,\perp}$ no longer represents the physical transverse momentum and does not directly influence the electron energy. As illustrated in figure~\ref{fig:Fermi-surface}, the correct Fermi surface for electrons in a given $n$-th  Landau level is characterized by a pair of longitudinal momenta that satisfy the equation $\sqrt{2n|eB|+p_{z,F}^2+m_e^2} = \mu_e$.

It is worth noting that the electron mass appears in the expression for the neutrino-number production rate in Eq.~(\ref{rate-02}) only through the electron's energy $E_{e,n}$. However, since the rate is primarily determined by the states near the Fermi surface and the electron mass is much smaller than the chemical potential ($m_e\ll \mu_e$), its effect is negligible. Therefore, we can safely neglect it in our calculation below.

Recognizing that the dominant contribution to neutrino emission arises from the processes with quarks within narrow energy bands ($\sim T$) near their Fermi surfaces, the expression for the rate in Eq.~(\ref{rate-02}) can be simplified by replacing the quark momenta with their Fermi momenta at all places in the integrand, except within the distribution functions. Furthermore, given that typical neutrino momenta are on the order of the temperature, which is small compared to the electron pseudo-momentum (recall that $p_e \geq v_F\mu_e$), it is justified to approximate $\bm{q}=\bm{p}_{e}-\bm{p}_{\nu}\approx \bm{p}_{e}$. Under these assumptions, we derive 
\begin{eqnarray}
&&\frac{\partial f_\nu(t,\bm{p}_\nu)}{\partial t}  \nonumber\\
&&= \frac{N_c  G_F^2\cos^2\theta_C}{ 2  \pi^3}v_F \mu_u \mu_d
\sum_{n=0}^{\infty} (-1)^n
\int \frac{dp_{e,z} d(p_{e,\perp}^2) e^{-p_{e,\perp}^2\ell^2} }{p_e} 
 \Theta(p_e) \int dk \,
n_F(p_{\nu}-E_k-E_{e,n}+\mu_d)\nonumber\\
&& \times  n_F(E_k-\mu_u)  n_F(E_{e,n}-\mu_e) \Bigg\{\left[ 1 - \left(1+ v_F \frac{\mu_u}{p_e} \cos\theta_{eu}\right) \frac{p_{e,z}p_{\nu,z} }{\mu_d p_{\nu}} \right]
\left( \frac{Y_{e,0}}{E_{e,n}}  -v_F \frac{(\bm{Y}_{e}\cdot \bm{p}_{e})}{p_e E_{e,n}} \cos\theta_{eu} \right) \nonumber\\
&& + v_F^2 \frac{1-\cos^2\theta_{eu}}{2}\frac{\mu_u p_{\nu,z} }{\mu_d p_{\nu}}
\left( \frac{Y_{e,z}}{E_{e,n}} -\frac{p_{e,z} (\bm{Y}_{e}\cdot \bm{p}_{e})}{ p_e^2 E_{e,n}}\right) 
\Bigg\}, 
\label{rate-04}
\end{eqnarray}
where we additionally performed the integration over the azimuthal angle $\phi_e$, determining the direction of the electron pseudo-momentum $\bm{p}_{e,\perp}$ in the plane perpendicular to the magnetic field. Note that scalar product $(\bm{Y}_{e}\cdot \bm{p}_{e})$, remaining in Eq.~(\ref{rate-04}), is independent of the angular coordinate $\phi_e$ and takes the following explicit form:
\begin{eqnarray}
(\bm{Y}_{e}\cdot \bm{p}_{e}) &=& p_{e,z}^2 \left[L_{n}\left(2 p_{e,\perp}^2\ell^2\right) -L_{n-1}\left(2 p_{e,\perp}^2\ell^2\right)  \right]
-s_\perp E_{e,n} p_{e,z}\left[L_{n}\left(2 p_{e,\perp}^2\ell^2\right) +L_{n-1}\left(2 p_{e,\perp}^2\ell^2\right)  \right]\nonumber\\
&& -4 p_{e,\perp}^2 L_{n-1}^1\left(2 p_{e,\perp}^2\ell^2\right).
\label{Ye-Pe}
\end{eqnarray}
In both Eqs.~(\ref{rate-04}) and (\ref{Ye-Pe}), we kept only the contribution of electron states with positive energies ($\lambda=1$) and dropped the contribution of positrons ($\lambda=-1$), which are negligible in dense quark matter with a relatively large $\mu_e$ compared to the temperature.

Finally, dropping the terms odd in $p_{e,z}$ and integrating over the up quark momentum $k$ using the table integral in Eq.~(\ref{int_01app}), we obtain the final expression for the neutrino-number production rate
{\small \begin{eqnarray}
&&\frac{\partial f_\nu(t,\bm{p}_\nu)}{\partial t}  
= \frac{N_c  G_F^2\cos^2\theta_C}{2 \pi^3}v_F \mu_u \mu_d
\sum_{n=0}^{\infty} (-1)^n
\int dp_{e,z} d(p_{e,\perp}^2) e^{-p_{e,\perp}^2\ell^2}  \Theta(p_e) 
n_B(p_{\nu}-E_{e,n}+\mu_e)  
 \nonumber\\
&&\times  \frac{p_{\nu}-E_{e,n}+\mu_e}{p_e} n_F(E_{e,n}-\mu_e) \Bigg\{ \left[1
+ \frac{ E_{e,n}}{2 \mu_u } \left(1-\frac{v_F^2 (\mu_d^2-\mu_u^2)}{p_e^2}\right) \right]
\left[L_{n}\left(2 p_{e,\perp}^2\ell^2\right) -L_{n-1}\left(2 p_{e,\perp}^2\ell^2\right)  \right]
 \nonumber\\
&&+\frac{s_\perp p_{\nu,z} }{4 \mu_u \mu_d p_{\nu} } \Bigg[ p_{e,z}^2 
\left(1+ \frac{v_F^2 (\mu_d^2-\mu_u^2)}{p_e^2} \right) 
\left(1-\frac{v_F^2 (\mu_d^2-\mu_u^2)}{p_e^2} +\frac{2\mu_u }{E_{e,n}} \right) 
  \nonumber\\
&& + \frac{p_{e,\perp}^2 }{2}
 \left(1-\frac{v_F^2(\mu_d+\mu_u)^2}{p_e^2}\right)
  \left(1-\frac{v_F^2\mu_e^2}{p_e^2}\right) \Bigg]
\left[L_{n}\left(2 p_{e,\perp}^2\ell^2\right) -L_{n-1}\left(2 p_{e,\perp}^2\ell^2\right)  \right]
\Bigg\}.
\label{rate-07}
\end{eqnarray}}
In the derivation, we used the explicit expression for $ \cos\theta_{eu}$, given in Eq.~\eqref{cos-theta}, and took into account the following identity for the Laguerre polynomials:  $z L_{n-1}^1(z) = -n\left[L_{n}(z)-L_{n-1}(z)\right]$. 

\subsection{Neutrino energy and net longitudinal momentum emission rates}

The neutrino-number production rate in Eq.~(\ref{rate-07}) is one of our main results. In this subsection, we utilize it to determine the energy and net longitudinal momentum emission rates for strongly magnetized dense quark matter. The corresponding rates are defined as follows:
\begin{eqnarray}
\dot{\cal E}_\nu &=& 2 \int \frac{d^3\bm{p}_{\nu}}{(2\pi)^3} p_{\nu,0} \frac{\partial f_\nu(t,\bm{p}_\nu)}{\partial t},\\
\dot{\cal P}_{\nu,z} &=& 2 \int \frac{d^3\bm{p}_{\nu}}{(2\pi)^3} p_{\nu,z} \frac{\partial f_\nu(t,\bm{p}_\nu)}{\partial t}.
\end{eqnarray}
Note that we included an additional factor of $2$ to account for the combined rate from both Urca processes, $u+ e^- \rightarrow d+\nu_e$ and $d\rightarrow u +e^- + \bar{\nu}_e$, one of which produces neutrinos and the other antineutrinos.

A nonzero net longitudinal momentum emission rate, $\dot{\cal P}_{\nu,z}$, indicates that neutrinos are emitted asymmetrically relative to the magnetic field direction and its opposite. Such an asymmetry is indeed anticipated in magnetized quark matter, where the combined effects of spin magnetization and parity violation generate a spatial imbalance in neutrino emission \cite{Sagert:2007as,Ayala:2018kie,Ayala:2024wgb}. 

Taking into account the explicit expression in Eq.~(\ref{rate-07}), we obtain
{\small\begin{eqnarray}
\dot{\cal E}_\nu &=& \frac{12 N_c G_F^2\cos^2\theta_C T^5}{\pi^5} v_F \mu_u \mu_d
\sum_{n=0}^{\infty} \frac{(-1)^n}{\ell^2}
\int_0^{\infty} \int_0^{\infty} \frac{\Theta(u,v) du dv}{\sqrt{u}\sqrt{u+v}} \frac{ e^{-v} }{e^{\epsilon_n^u}+1}
\left( \mbox{Li}_{5}\left(e^{\epsilon_n^u} \right) -\frac{\epsilon_n^u}{4}\mbox{Li}_{4}\left(e^{\epsilon_n^u} \right)\right) 
\nonumber\\
& \times & 
 \left[L_{n}\left(2 v\right) -L_{n-1}\left(2 v\right)  \right]
\left[ 1 + \frac{ \sqrt{2n+u} }{2 \ell \mu_u} 
\left(1- \frac{v_F^2 \mu_e(\mu_d+\mu_u)\ell^2}{u+v}\right) \right]  , \label{rate-energy-a}
\end{eqnarray}}
and  
{\small \begin{eqnarray}
\dot{\cal P}_{\nu,z} &=& s_\perp \frac{N_c  G_F^2\cos^2\theta_C T^5}{\pi^5} v_F
\sum_{n=0}^{\infty} \frac{(-1)^n}{\ell^4}
\int_0^{\infty} \int_0^{\infty} \frac{\Theta(u,v) du dv}{\sqrt{u}\sqrt{u+v}} \frac{ e^{-v} }{e^{\epsilon_n^u}+1}
\left( \mbox{Li}_{5}\left(e^{\epsilon_n^u} \right) -\frac{\epsilon_n^u}{4}\mbox{Li}_{4}\left(e^{\epsilon_n^u} \right)\right)  
\nonumber\\
& \times & 
\left[L_{n}\left(2 v\right) + L_{n-1}\left(2 v\right)  \right] 
\left[
u\left(1+ \frac{v_F^2 \mu_e(\mu_d+\mu_u)\ell^2}{u+v} \right) 
\left(1 - \frac{v_F^2 \mu_e(\mu_d+\mu_u)\ell^2}{u+v} +\frac{2 \ell \mu_u}{\sqrt{2n+u} } \right) \right. \nonumber\\
&& \left.
+ \frac{v}{2}\left(1- \frac{v_F^2 (\mu_d+\mu_u)^2\ell^2}{u+v}\right)\left(1- \frac{v_F^2 \mu_e^2\ell^2}{u+v}\right) 
\right] ,  
\label{rate-momentum-a}
\end{eqnarray}}
respectively. In deriving these results, we calculated the integral over the neutrino three-momentum $\bm{p}_{\nu}$ by using the table integral in Eq.~(\ref{int_05app}). Additionally, we introduced new dimensionless integration variables $u = p_{e,z}^2\ell^2$  and  $v = p_{e,\perp}^2\ell^2$, as well as adopted the following shorthand notations:
\begin{eqnarray}
\epsilon_n^u&=&\frac{\sqrt{2n+u}-\mu_e\ell}{T\ell} , \\
\Theta(u,v) &\equiv &
 \theta\left(u+v-v_F^2\mu_e^2\ell^2\right)\theta\left[v_F^2(\mu_d+\mu_u)^2\ell^2-u-v\right].
 \end{eqnarray}
Recall that the electron mass is negligible compared to the electron chemical potential ($m_e\ll\mu_e$) and thus can be safely ignored in the calculation.

\subsection{Simplified approximation in the limit $T\to 0$}

The energy and momentum emission rates in Eqs.~(\ref{rate-energy-a}) and (\ref{rate-momentum-a}) can be further simplified in the limit of extremely low temperature. Quantitatively, this regime is reached when temperatures become much lower than the energy spacing between adjacent Landau levels at the electron Fermi surface, i.e., $T\ll |eB|/\mu_e$, assuming $\sqrt{|eB|}\lesssim \mu_e$. (For stronger fields, with $\sqrt{|eB|}\gtrsim \mu_e$ and only the lowest Landau level occupied, the same condition holds when $T\ll \sqrt{|eB|}$.)

In this case, one may assume that the longitudinal momenta of electrons are restricted to their values at the Landau-level dependent Fermi surfaces, defined as  $p^{(\pm)}_{z,F}=\pm \sqrt{\mu_e^2-2n|eB|}$, provided $n\leq \mu_e^2/(2|eB|)$, see the schematic illustration in figure~\ref{fig:Fermi-surface}. Then, the integration over variable $u$ in Eqs.~(\ref{rate-energy-a}) and (\ref{rate-momentum-a}) can be easily evaluated using the table integral in Eq.~(\ref{int_04app}). The resulting expressions are
\begin{eqnarray}
 \dot{\cal E}^{(0)}_\nu &=& \frac{457 \pi N_c  G_F^2\cos^2\theta_C}{5040} v_F \mu_u \mu_d\mu_e T^6 \left( 1+ \frac{\mu_e}{2\mu_u} \right)
 \nonumber\\
 &&\times  
 \sum_{n=0}^{n_{\rm max}} \frac{(-1)^n}{\sqrt{u_n}}
 \int \frac{\Theta(u_n,v) e^{-v}  dv  }{\sqrt{u_n+v}}  
 \left( 1 - \frac{v_F^2 \mu_e^2\ell^2}{u_n+v}  \right)
 \left[L_{n}\left(2 v\right) -L_{n-1}\left(2 v\right)  \right] ,
 \label{rate-energy-0}
\end{eqnarray}
and
\begin{eqnarray}
 \dot{\cal P}^{(0)}_{\nu,z} &=& s_\perp\frac{457 \pi N_c  G_F^2\cos^2\theta_C}{60480\ell^2} v_F \mu_e T^6
 \sum_{n=0}^{n_{\rm max}} \frac{(-1)^n}{\sqrt{u_n}}
 \int  \frac{\Theta(u_n,v) e^{-v}  dv  }{\sqrt{u_n+v}} 
 \left( 1 - \frac{v_F^2\mu_e^2\ell^2}{u_n+v}\right) \nonumber\\
 && \times 
 \left[ u_{n}  \left(1+\frac{v_F^2\mu_e(\mu_d+\mu_u)\ell^2}{u_n+v}\right) \frac{\mu_d+\mu_u}{\mu_{e}}
 +\frac{v}{2}\left(1-\frac{v_F^2(\mu_d+\mu_u)^2\ell^2}{u_n+v}\right)\right]  
 \left[L_{n}\left(2 v\right) + L_{n-1}\left(2 v\right)  \right]  ,\nonumber\\
 &&
 \label{rate-momentum-0}
\end{eqnarray}
where $u_{n}\equiv \mu_e^2\ell^2 -2n$ and $n_{\rm max}$ denotes the highest occupied Landau level and is given by the integer part of $\mu_e^2/(2|eB|)$. 

Note that both rates in Eqs.~(\ref{rate-energy-0}) and (\ref{rate-momentum-0}) formally diverge when $u_n=0$, corresponding to the condition where the Fermi energy matches the $n$-th Landau level threshold, i.e., $\sqrt{2n|eB|} =\mu_e$. However, these divergences are artifacts of the approximation and are unphysical at any finite temperature. Indeed, it can be verified that such divergences do not appear in the original expressions in Eqs.~(\ref{rate-energy-a}) and (\ref{rate-momentum-a}) when $T\neq 0$. This conclusion will be also confirmed by our numerical results in the next section.

\subsection{Strong magnetic field limit}

Before concluding this section, let us briefly examine, for completeness, the case of an ultra-strong magnetic field where electrons are restricted to the lowest Landau level (LLL) states. This regime is formally reached when $|eB|\gtrsim \mu_e^2$. Utilizing the simplest approximation for the rates given by Eqs.~(\ref{rate-energy-0}) and (\ref{rate-momentum-0}), we derive 
\begin{eqnarray}
 \dot{\cal E}^{\rm (LLL)}_\nu &\simeq& \frac{457 \pi^{3/2} N_c  G_F^2\cos^2\theta_C}{5040\ell } v_F \mu_u \mu_d  T^6\left( 1 + \frac{\mu_e}{2\mu_u} \right)
\left( \left(1+2v_F^2\mu_e^2\ell^2 \right)e^{\mu_e^2\ell^2} \mbox{erfc} \left(\mu_e\ell\right) -\frac{2v_F^2\mu_e\ell}{\sqrt{\pi}}  \right) ,\nonumber\\
 &&
 \label{rate-energy-LLL}
\end{eqnarray}
and
\begin{eqnarray}
 \dot{\cal P}^{\rm (LLL)}_{\nu,z} &\simeq & s_\perp \frac{457 \pi^{3/2} N_c  G_F^2\cos^2\theta_C}{60480\ell } v_F (\mu_u+\mu_d)^2 T^6 
 \Bigg[ \frac{v_F^2}{\sqrt{\pi}} \mu_e \ell (3+2v_F^2 \mu_e^2\ell^2)
 -\frac{2v_F^2\mu_e^2\ell }{\sqrt{\pi} (\mu_u+\mu_d) } \nonumber\\
&&  +\frac{\mu_e(1+2v_F^2 \mu_e^2\ell^2)}{2\sqrt{\pi} (\mu_u+\mu_d)^2 \ell}
- e^{\mu_e^2\ell^2} \mbox{erfc} \left(\mu_e\ell\right)\left(
\frac{v_F^2}{2}  \left(1+2(3+v_F^2)\mu_e^2\ell^2+4v_F^2\mu_e^4\ell^4\right) \right. \nonumber \\
&&\left.
-\frac{\mu_e\left(1+2v_F^2\mu_e^2\ell^2\right)}{\mu_u+\mu_d}
-\frac{1-2(1+v_F^2)\mu_e^2\ell^2-4v_F^2\mu_e^4\ell^4}{4(\mu_u+\mu_d)^2\ell^2} \right)
\Bigg], 
 \label{rate-momentum-LLL}
\end{eqnarray}
where $\mbox{erfc}(z)$ is the complementary error function. In the derivation we assumed that $ \mu_e^2\lesssim |eB|\ll (\mu_d+\mu_u)^2$. As we will demonstrate in the next section, the numerical results indeed approach the rates dominated by the LLL electrons in the regime of extremely strong magnetic fields. Note, however, that the momentum rate (\ref{rate-momentum-LLL}) may be less accurate than the energy rate (\ref{rate-energy-LLL}) at nonzero temperatures. This is partly due to the more complicated dependence of the momentum emission rate on quark matter model parameters, including the quark chemical potentials. Nevertheless, in the limit $T\to 0$, the numerical results tend to approach closely these LLL approximations, as shown in figure~\ref{fig.rates_20-600-T1mue40} below.

It is instructive to note that the rate in Eq.~(\ref{rate-energy-LLL}) is an increasing function of the magnetic field. From a physics viewpoint, there are at least two main reasons for the enhanced neutrino emission in the LLL approximation. First, the rate increases because the overall degeneracy of the electron Landau levels grows with field strength. Second, the usual collinear restrictions on particle momenta are maximally lifted for the LLL electrons. Both effects are reflected in Eq.~(\ref{rate-energy-LLL}). Compared to the zero-field scaling in Iwamoto's expression for the rate, see Eq.~(\ref{dot-E-app-Iwamoto}), the overall factor of $\mu_e$ is replaced by $1/\ell =\sqrt{|eB|}$ in the LLL approximation, and there is no suppression from the overall factor of $\alpha_s$. [Note also that the additional function in parenthesis is an increasing function of $|eB|/\mu_e^2$, with values that remain within a finite range of approximately $0.15$ to $1$.]

\section{Numerical results}
\label{sec:Numerics}

Having derived the analytical results in the preceding section, we now focus on the numerical investigation of the neutrino emission rates from magnetized quark matter. Our primary goal is to examine the general effects of a quantizing magnetic field rather than to model a specific compact star. Accordingly, we vary model parameters, such as the magnetic field and chemical potentials, independently. To ensure realistic conditions for quark-matter cores, we adopt the following assumptions: (i) the quark chemical potentials ($\mu_f$ where $f=u,d$) are around $300~\mbox{MeV}$ (or larger) and the electron chemical potential ($\mu_e$) is on the order of $50~\mbox{MeV}$; (ii) the temperature is sufficiently low, $T\lesssim 5~\mbox{MeV}$, so that neutrino trapping is absent; and (iii) while the background magnetic field is strong, it remains within a realistic range, i.e., $\sqrt{|eB|} \lesssim 25~\mbox{MeV}$, corresponding to field strengths below about $10^{17}~\mbox{G}$. Also, given the requirement of $\beta$-equilibrium, we will always enforce the following relation between chemical potentials: $\mu_d=\mu_u+\mu_e$. 

In order to focus on the role of a background magnetic field on the rates, it is instructive to consider the following dimensionless ratios
\begin{eqnarray}
\frac{\dot{\cal E}_\nu}{\dot{\cal E}_\nu (B=0)}  &=&  
\frac{60480 \mu_u }{457\pi^6 (1-v_F^2)  C_T \mu_e  (\mu_d+\mu_u) T}
\sum_{n=0}^{\infty} \frac{(-1)^n}{\ell^2}
\int_0^{\infty} \int_0^{\infty} \frac{\Theta(u,v) du dv}{\sqrt{u}\sqrt{u+v}} \frac{ e^{-v} }{e^{\epsilon_n^u}+1} \nonumber\\ 
&&\times
\left( \mbox{Li}_{5}\left(e^{\epsilon_n^u} \right) -\frac{\epsilon_n^u}{4}\mbox{Li}_{4}\left(e^{\epsilon_n^u} \right)\right) 
 \left[ 1 + \frac{ \sqrt{2n+u} }{2 \ell \mu_u} 
\left(1- \frac{v_F^2 \mu_e(\mu_d+\mu_u)\ell^2}{u+v}\right) \right]  
\nonumber\\
&& \times \left[L_{n}\left(2 v\right) -L_{n-1}\left(2 v\right)  \right],
\label{ratio-energy-a}
\end{eqnarray}
and
\begin{eqnarray}
\frac{\dot{\cal P}_{\nu,z}}{\dot{\cal E}_\nu (B=0)} 
&=& s_\perp  \frac{5040 }{457\pi^6 (1-v_F^2)  C_T\mu_e \mu_d(\mu_d+\mu_u)T} 
\sum_{n=0}^{\infty}  \frac{(-1)^n}{\ell^4}
\int_0^{\infty} \int_0^{\infty} \frac{\Theta(u,v) du dv}{\sqrt{u}\sqrt{u+v}} \frac{ e^{-v} }{e^{\epsilon_n^u}+1}\nonumber\\
&& \times
\left( \mbox{Li}_{5}\left(e^{\epsilon_n^u} \right) -\frac{\epsilon_n^u}{4}\mbox{Li}_{4}\left(e^{\epsilon_n^u} \right)\right)  \Bigg[
u\left(1+ \frac{v_F^2 \mu_e(\mu_d+\mu_u)\ell^2}{u+v} \right) 
\left(1 - \frac{v_F^2 \mu_e(\mu_d+\mu_u)\ell^2}{u+v} \right. \nonumber\\
&&\left. +\frac{2 \ell \mu_u}{\sqrt{2n+u} } \right) + \frac{v}{2}\left(1- \frac{v_F^2 (\mu_d+\mu_u)^2\ell^2}{u+v}\right)\left(1- \frac{v_F^2 \mu_e^2\ell^2}{u+v}\right) 
\Bigg] 
\left[L_{n}\left(2 v\right) + L_{n-1}\left(2 v\right)  \right] ,\nonumber\\
&& 
\label{ratio-momentum-a}
\end{eqnarray}
instead of calculating the rates themselves, as defined in Eqs.~(\ref{rate-energy-a}) and (\ref{rate-momentum-a}). We will also employ similar ratios when analyzing the expressions in Eqs.~(\ref{rate-energy-0}) and (\ref{rate-momentum-0}). It is important to note that the rates are normalized by the zero-field rate given in Eq.~(\ref{dot-E-CT-app-B0}). Compared to the result in Eq.~(\ref{dot-E-app-B0}), this expression includes an additional factor $C_T$, which reflects a more accurate treatment of the integration over the electron momentum, without artificially enforcing $p_e\approx \mu_e$ in the integrand. This definition is particularly appropriate as it closely aligns with the approximations employed in deriving our main results in Eqs.~(\ref{rate-energy-a}) and (\ref{rate-momentum-a}) in the presence of a background magnetic field.

\subsection{Rates at very low temperatures}

\begin{figure}[t]
\centering
  \subfigure[]{\includegraphics[width=0.45\textwidth]{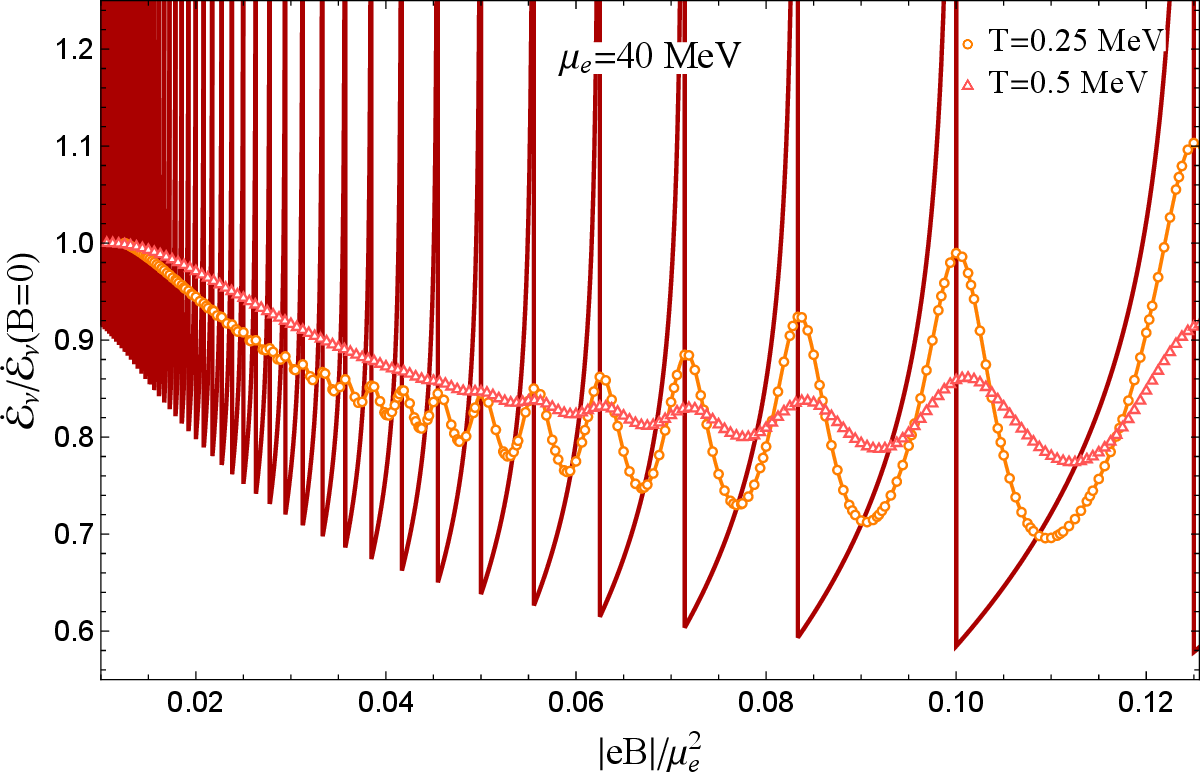}}
  \hspace{0.05\textwidth}
  \subfigure[]{\includegraphics[width=0.45\textwidth]{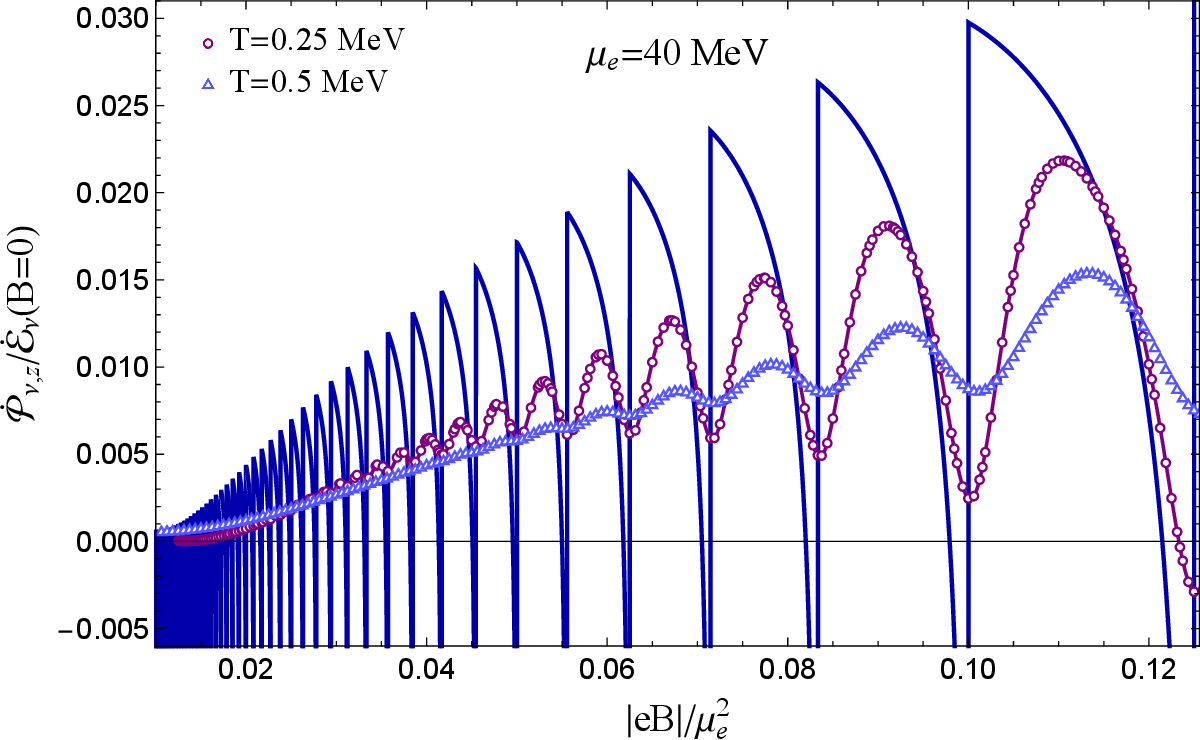}}
\caption{The neutrino energy (a) and momentum (b) emission rates in the low-temperature approximation (solid lines),  see Eqs.~(\ref{rate-energy-0}) and (\ref{rate-momentum-0}), and the exact finite-temperature results from Eqs.~(\ref{rate-energy-a}) and (\ref{rate-momentum-a}) for two fixed temperatures: $T=0.25~\mbox{MeV}$ (empty circles) and $T=0.5~\mbox{MeV}$ (empty triangles).}
\label{fig.rates-mue40-T0}
\end{figure}

We begin by analyzing the neutrino energy and  longitudinal momentum emission rates given in Eqs.~(\ref{rate-energy-0}) and (\ref{rate-momentum-0}), which are derived using the simplest approximation that is expected to be accurate only in the limit of extremely low temperatures. The corresponding numerical results are represented by solid lines in figure~\ref{fig.rates-mue40-T0}, spanning a magnetic field range from approximately $|eB|\simeq 20~\mbox{MeV}^2$ to $|eB|\simeq 200~\mbox{MeV}^2$. For illustrative purposes, we used the following default values for the chemical potentials: $\mu_u = 260~\mbox{MeV}$ and $\mu_e = 40~\mbox{MeV}$. It is important to note, however, that the results, when normalized by the zero-field rate as displayed in figure~\ref{fig.rates-mue40-T0}, are expected to be largely independent of the specific values of the quark chemical potentials. To account for the quark Fermi-liquid effects, we use the strong coupling constant $\alpha_s=0.3$. 

As predicted, the rates exhibit a characteristic sawtooth dependence on the dimensionless parameter $|eB|/\mu_e^2$, with divergences occurring at points where the Fermi energy aligns with the thresholds of individual Landau levels, i.e., $|eB|/\mu_e^2=1/(2n)$ for all positive integer $n$. We emphasize, however, that the divergences are artifacts of the approximation used. This conclusion is further supported by comparing the corresponding results with the correct numerical rates calculated from Eqs.~(\ref{ratio-energy-a}) and (\ref{ratio-momentum-a}) at two fixed temperatures, $T=0.25~\mbox{MeV}$ and $T=0.5~\mbox{MeV}$, also shown in figure~\ref{fig.rates-mue40-T0}. As seen, the oscillatory behavior of the numerical results becomes increasingly pronounced as the temperature decreases, gradually approaching the extreme sawtooth pattern as $T\to 0$.

It is worth noting that, while the neutrino energy emission rate  is always positive, the longitudinal momentum emission rate does not have a definite sign. This indicates that the net momentum emission is directed along the magnetic field for certain values of $|eB|/\mu_e^2$ and in the opposite direction for others. As we will elaborate in section~\ref{Partial-LL-contributions} below, this behavior arises because weak processes involving electrons near the equatorial regions of the Fermi surface (where $p_{e,z}\simeq 0$) predominantly emit neutrinos in the opposite direction compared to those involving electrons near the polar regions (where $|p_{e,z}|\simeq \mu_e$), see figure~\ref{fig:Fermi-surface}.

\subsection{Magnetic field dependence of rates at finite temperatures}

\begin{figure}[t]
\centering
  \subfigure[]{\includegraphics[width=0.45\textwidth]{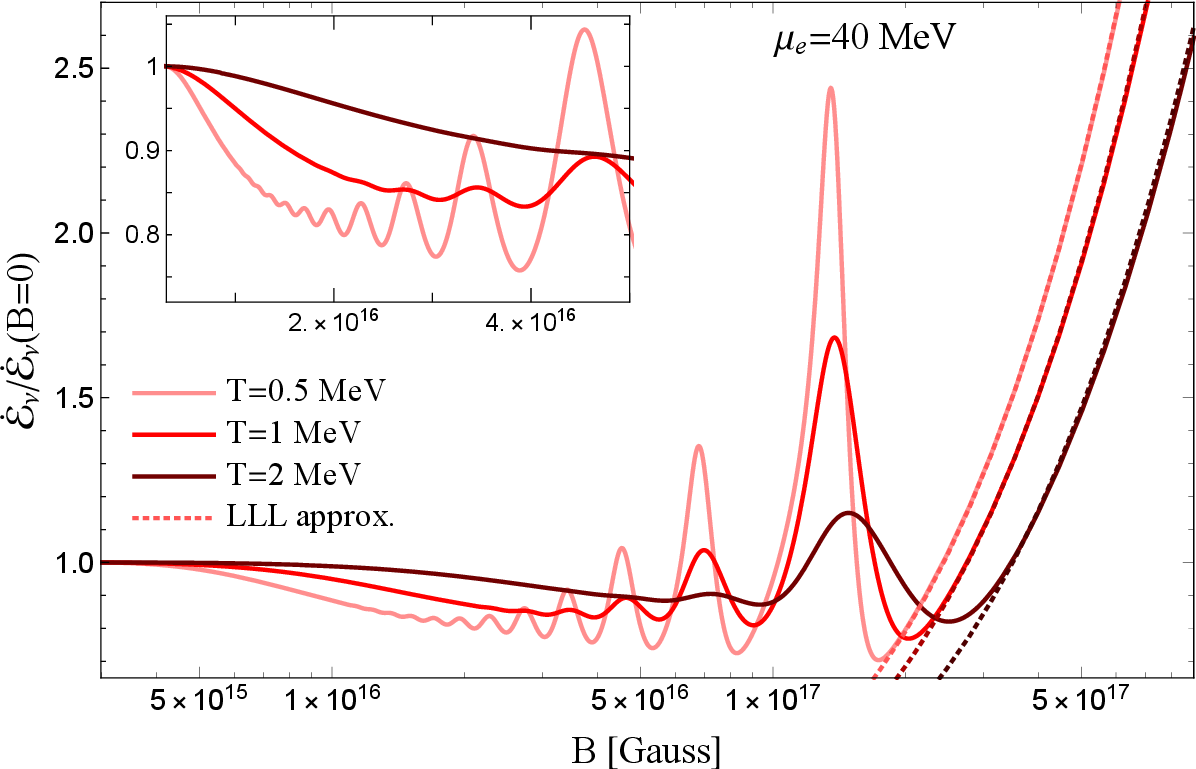}}
  \hspace{0.05\textwidth}
  \subfigure[]{\includegraphics[width=0.45\textwidth]{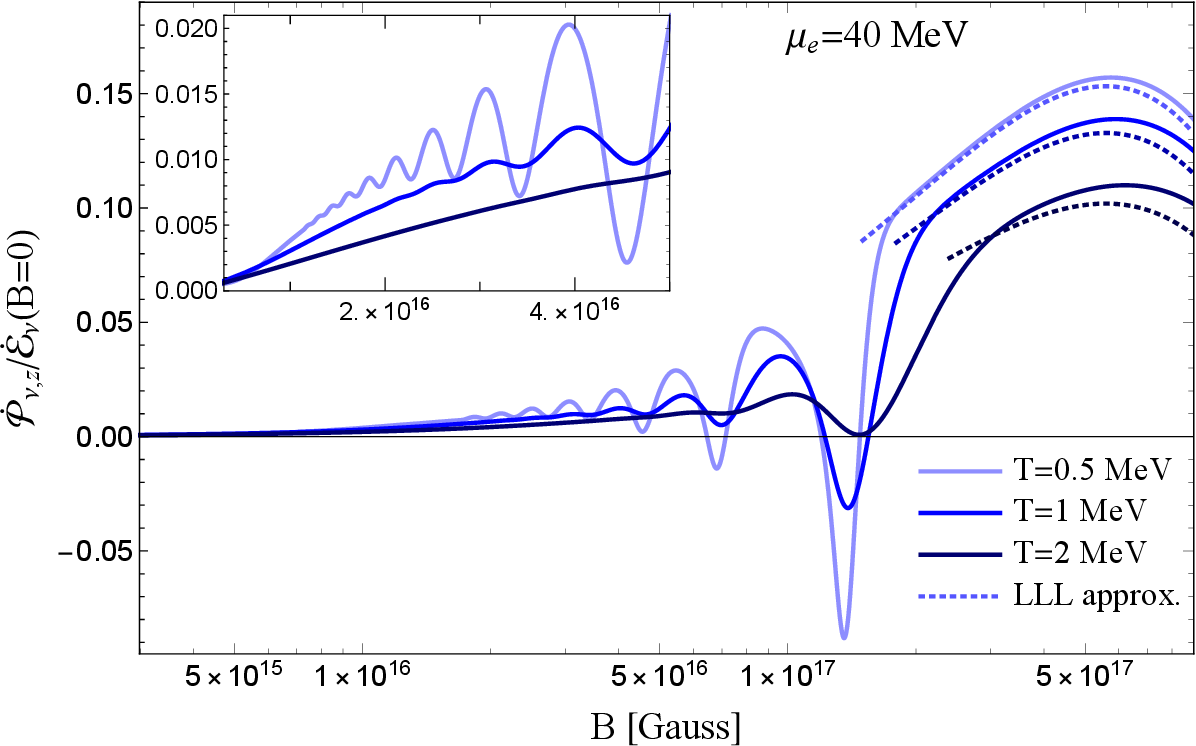}}
\caption{The neutrino energy (a) and momentum (b) emission rates as functions of the magnetic field strength for three fixed temperatures: $T=0.5~\mbox{MeV}$, $T=1~\mbox{MeV}$, and $T=2~\mbox{MeV}$. The rates in the LLL approximation are represented by dotted lines. The insets show a close-up view of the regions with the magnetic field strength below $5\times 10^{16}~\mbox{G}$.}
\label{fig.rates_20-600-T1mue40}
\end{figure}

Our main numerical results for the rates given by Eqs.~(\ref{ratio-energy-a}) and (\ref{ratio-momentum-a}) as functions of the magnetic field are presented in figure~\ref{fig.rates_20-600-T1mue40} for three fixed temperatures: $T=0.5~\mbox{MeV}$, $T=1~\mbox{MeV}$, and $T=2~\mbox{MeV}$.  As before, we used the default value $\mu_e=40~\mbox{MeV}$.  To represent a broad range of magnetic fields, from approximately $|eB|\simeq 18~\mbox{MeV}^2$ to $|eB|\simeq 5000~\mbox{MeV}^2$, which is equivalent to the field strengths in a range from about $B\simeq 3\times 10^{15}~\mbox{G}$ to $B\simeq 8.5\times 10^{17}~\mbox{G}$, we use a logarithmic scale for the horizontal axis. Note that both energy and momentum rates in figure~\ref{fig.rates_20-600-T1mue40} are normalized to the zero-field energy rate. To convert the squared energy units of $|eB|$ into gauss for the abscissa, we used the following conversion formula:
\begin{equation}
B=1.69\times 10^{14}\frac{|eB|}{\mbox{MeV}^2}~\mbox{G} .
\end{equation}
Since the energy rate is normalized to its zero-field value, all three dimensionless ratios must converge to $1$ as $B\to 0$. In practice, to calculate such data we first computed the ratio of the  $B\neq 0$ rates and the approximate rate at $B= 0$, given by Eq.~(\ref{dot-E-app-B0}) rather than the correct expression in Eq.~(\ref{dot-E-CT-app-B0}) that contains an extra factor of $C_T\neq 1$. Then the data point for the smallest magnetic field value (approximately $|eB|=20~\mbox{MeV}^2$) was used to extract the numerical factor $C_{T}$. For the three temperatures shown, the respective constants were $C_{T}(0.5~\mbox{MeV})\approx 1.2$, $C_{T}(1~\mbox{MeV})\approx 1.385$, and $C_{T}(2~\mbox{MeV})\approx 1.805$. The analogous numerical constant for the lowest temperature case used in figure~\ref{fig.rates-mue40-T0} was $C_{T}(0.25~\mbox{MeV})\approx 1.11$. While these constants are not identical, they are close to the zero-field values of the function $C_{T}$, which is approximated by the simple fit in Eq.~(\ref{CT-app-B0}) as a function of the dimensionless ratio $T/\mu_e$. Such numerical consistency provides additional verification that the evaluation was performed correctly.

To better understand the behavior of the rates at moderate magnetic fields, in figure~\ref{fig.rates_20-600-T1mue40} we added insets with a close-up view of the regions with the field strength below $5\times 10^{16}~\mbox{G}$. Unlike the main plots, which use a logarithmic scale for the magnetic field, the insets use a linear scale. As one can see, when the oscillatory behavior of the rates is disregarded in the whole region below about $10^{17}~\mbox{G}$, the average neutrino energy emission rate appears to decrease with increasing magnetic field strength, whereas the average momentum emission rate exhibits an upward trend. However, these trends become less pronounced as the temperature increases. The qualitative behavior changes at extremely large magnetic fields when the rates are dominated by the LLL electrons. 

Upon a close examination of the numerical data, we observe that the LLL contribution to the energy emission rate exhibits a monotonically increasing dependence on the field strength. In contrast, contributions from higher Landau levels combine to produce an oscillatory behavior with a progressively decreasing average value. At sufficiently large field strengths, the LLL contribution dominates, as contributions from higher levels become negligible.
For the momentum emission rate, the LLL contribution initially grows with increasing field strength, reaching a peak near $B \sim 5.7\times 10^{17}~\mbox{G}$, after which it decreases. In comparison, the collective contributions from higher Landau levels introduce oscillatory behavior with predominantly negative values.
 
The results shown in figure~\ref{fig.rates_20-600-T1mue40} reconfirm that both energy and momentum emission rates exhibit an oscillatory dependence on the magnetic field strength. As expected, local peaks in the energy emission rates occur at or slightly above the threshold values $|eB|=\mu_e^2/(2n)$ for all temperatures. However, the oscillation amplitude decreases with increasing temperature, and the peaks remain well-resolved only for sufficiently strong magnetic fields. In contrast, at sufficiently low values of $B$, the thermal effects tend to completely wash away oscillations. Interestingly, at the same threshold locations, the momentum emission rates exhibit local minima rather than maxima, while their response to thermal effects qualitatively mirrors that of the energy emission rates. From our numerical results, we find that the corresponding maxima (minima) remain resolved only when the Landau-level spacing at the Fermi surface, $|eB|/\mu_e$, exceeds the thermal energy scale, $\pi T$, or equivalently, when $|eB|\gtrsim \pi T \mu_e$.

At extremely high magnetic fields, $|eB|\gtrsim \mu_e^2$, where electrons primarily occupy the LLL, the energy and momentum emission rates exhibit intriguing behavior. Note that the rates in the LLL approximation, as given in Eqs.~(\ref{rate-energy-LLL}) and (\ref{rate-momentum-LLL}), are represented by dotted lines in figure~\ref{fig.rates_20-600-T1mue40}. In this regime, the neutrino energy emission rate increases rapidly with the magnetic field strength. Over a broad range, approximately from  $|eB|\simeq 1.2\times 10^3~\mbox{MeV}^2$ to $|eB|\simeq 6\times 10^3~\mbox{MeV}^2$ or, equivalently, from $B\simeq 2\times 10^{17}~\mbox{G}$ to $B\simeq 10^{18}~\mbox{G}$, the growth is nearly linear, with the emission rate reaching values almost five times the zero-field result. At even higher magnetic fields, the growth persists, but the dependence slowly transitions to a square-root behavior.

The momentum emission rate, in contrast, behaves differently at very high magnetic fields.  It initially increases, reaching a peak value of approximately $0.15$ around $B\simeq 5.7\times 10^{17}$, before decreasing as the field strength continues to rise. Remarkably, at approximately $|eB|\simeq 10^4~\mbox{MeV}^2$, the momentum rate crosses zero and becomes negative. (Formally, it becomes positive again for $|eB|\gtrsim 3.2\times 10^5~\mbox{MeV}^2$.) However, at such extreme field strengths ($|eB|\gtrsim 10^4~\mbox{MeV}^2$), the validity of the underlying approximations becomes questionable. Therefore, all results in this regime should be interpreted with caution.

\subsection{Partial contributions of individual Landau levels to the rates}
\label{Partial-LL-contributions}

\begin{figure}[t]
\centering
{\includegraphics[width=0.45\textwidth]{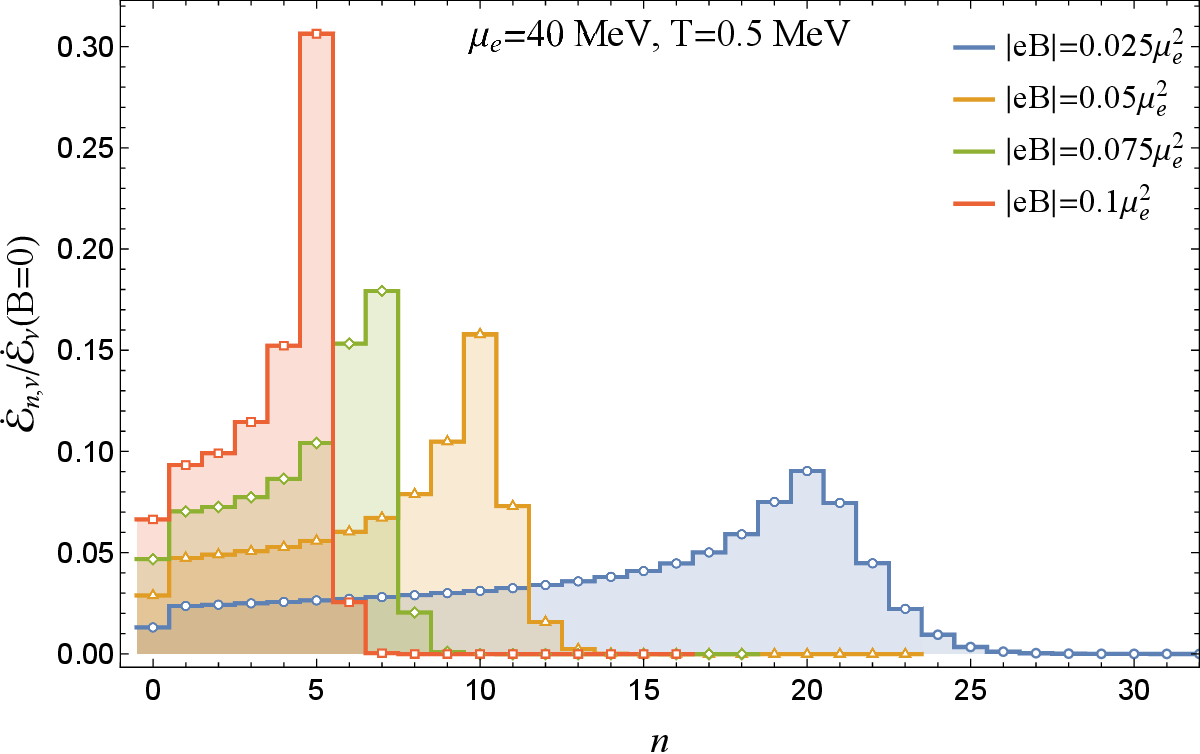}}
  \hspace{0.05\textwidth}
{\includegraphics[width=0.45\textwidth]{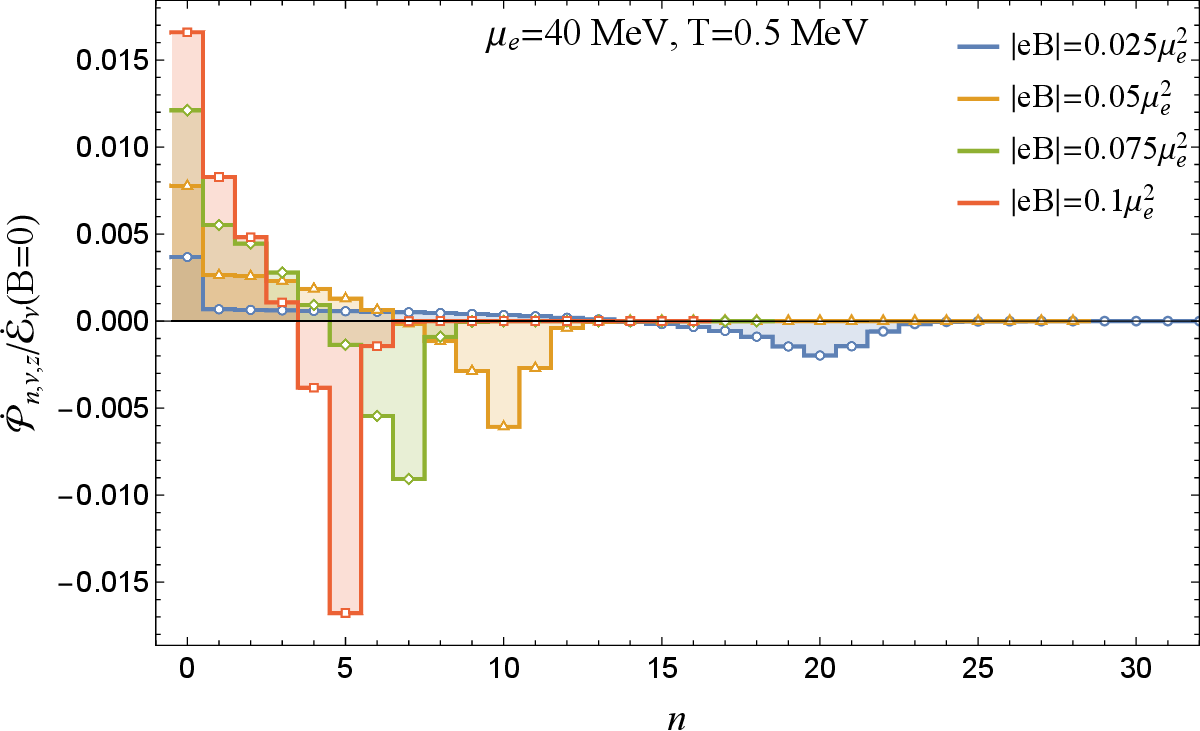}}\\
{\includegraphics[width=0.45\textwidth]{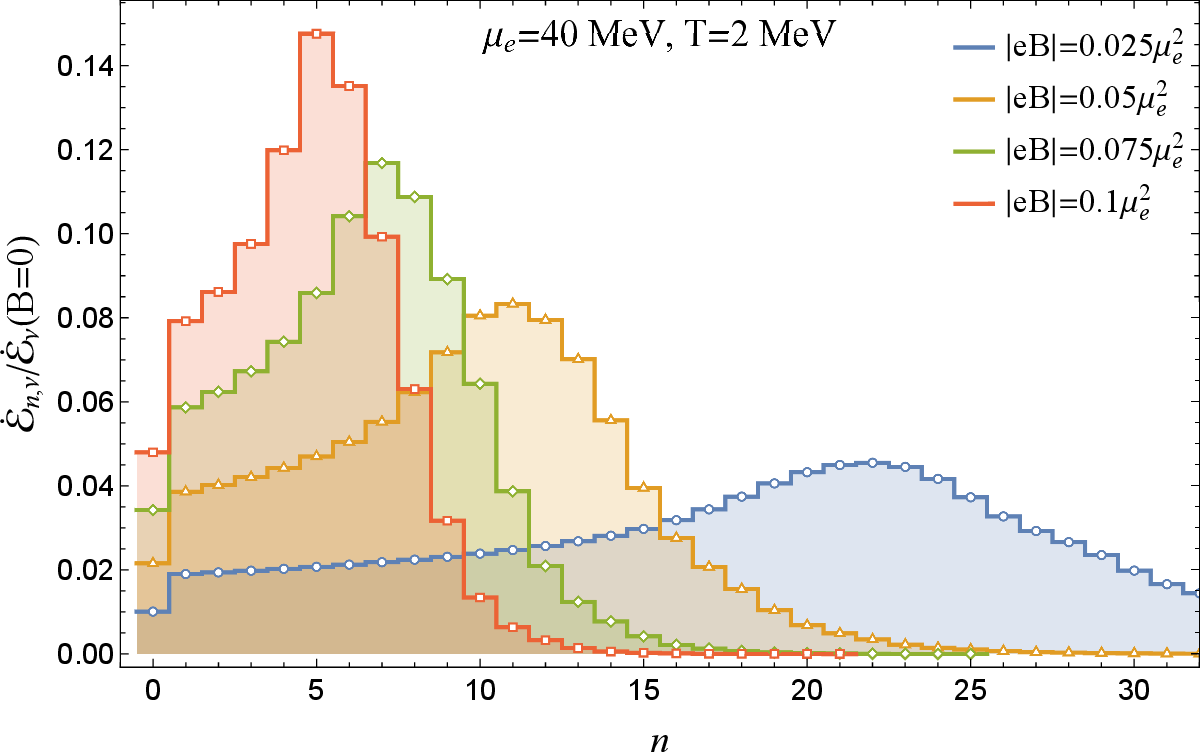}}
  \hspace{0.05\textwidth}
{\includegraphics[width=0.45\textwidth]{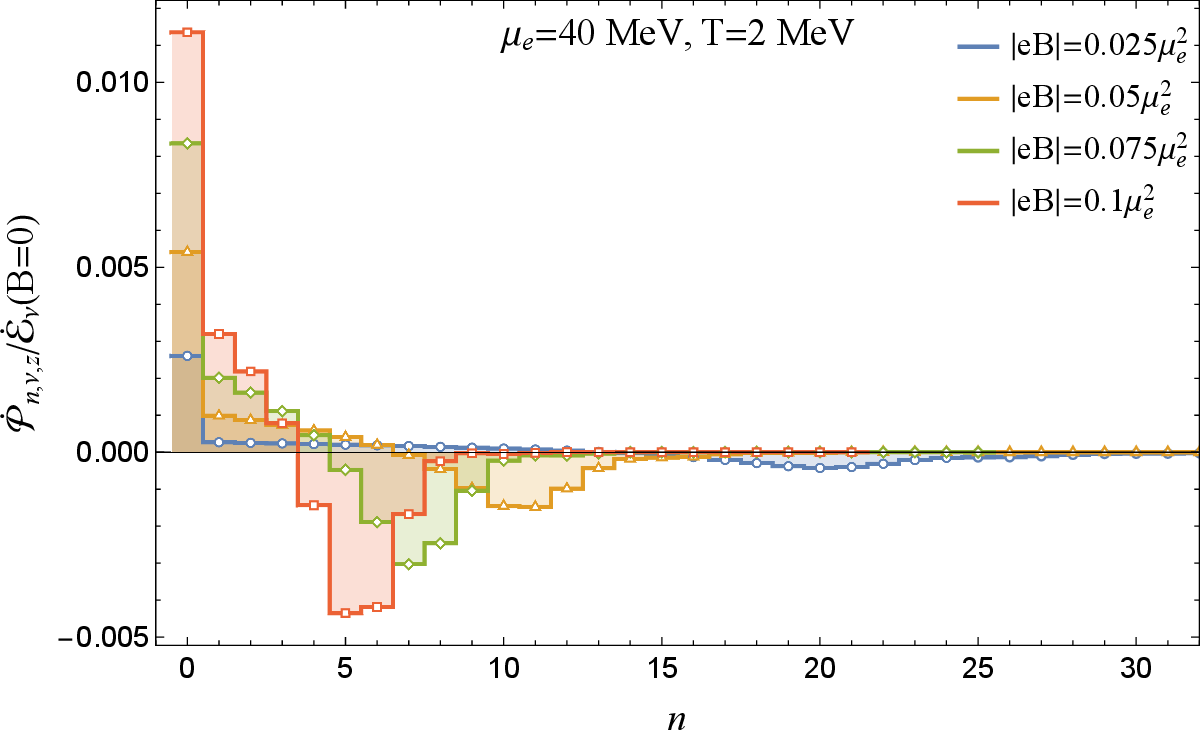}}
\caption{Partial contributions of Landau levels to the neutrino energy (left panels) and momentum (right panels) emission rate for several choices of the magnetic field and two different temperatures: $T=0.5~\mbox{MeV}$ (top panels) and $T=2~\mbox{MeV}$ (bottom panels).}
\label{fig.rates_LLs-mue40}
\end{figure}

To gain deeper insight into neutrino emission from magnetized dense quark matter, it is beneficial to examine the contributions of individual Landau levels to the relevant Urca processes. From a technical viewpoint, separating the partial contributions of different Landau levels is straightforward using Eqs.~(\ref{ratio-energy-a}) and (\ref{ratio-momentum-a}), which explicitly include sums over the Landau index $n$. Representative data for two fixed temperatures, $T=0.5~\mbox{MeV}$ and $T=2~\mbox{MeV}$, are presented in figure~\ref{fig.rates_LLs-mue40}, with the two top panels corresponding to the lower temperature and the two bottom panels to the higher temperature. Each panel displays four histograms corresponding to different values of the magnetic field: $|eB|/\mu_e^2=0.025$ (blue), $|eB|/\mu_e^2=0.05$ (orange), $|eB|/\mu_e^2=0.075$ (green), and $|eB|/\mu_e^2=0.1$ (red).

For energy emission, the largest partial contributions originate from electron states at the Fermi surface with the smallest values of $|p_{z,F}|$. These states are associated with the Landau level whose energy minimum lies closest to the chemical potential. The corresponding Landau-level index $n_{\rm max}$ is approximately given by the integer part of $\mu_e^2/(2|eB|)$, see figure~\ref{fig:Fermi-surface}. For the magnetic field values used in figure~\ref{fig.rates_LLs-mue40}, for example, the corresponding $n_{\rm max}$ values are $20$, $10$, $7$, and $5$. This finding is hardly surprising after revisiting the simplest approximation for the rates in Eqs.~(\ref{rate-energy-0}) and (\ref{rate-momentum-0}), where the contribution from the Landau level near the threshold diverges in the limit $T\to 0$.

Several additional features regarding the partial Landau-level contributions to the energy emission are worth noting. As seen from the two left panels in figure~\ref{fig.rates_LLs-mue40}, neighboring Landau levels generally give comparable contributions. However, the LLL contribution stands out as a clear outlier. Its contribution is suppressed by approximately a factor of two relative to the next level ($n=1$). This suppression likely arises from the fact that the LLL includes only spin-down states, unlike higher levels, which are twice as degenerate due to the presence of both spin-up and spin-down states. Another noteworthy observation is the significant contribution from Landau levels with $n> n_{\rm max}$, especially at moderately high temperatures and relatively weak magnetic fields. For instance, for $|eB|=0.025 \mu_e^2$ and $T=2~\mbox{MeV}$, more than a dozen Landau levels above the Fermi level contribute substantially to the emission rate.

The details of individual Landau-level contributions to the momentum emission rate, illustrated in the two right panels of figure~\ref{fig.rates_LLs-mue40}, are even more intriguing. Notably, the LLL  states near the Fermi surface play a distinct role again, but here they emerge as the single largest contributors to the net momentum emission. This behavior is likely to stem from the spin-down polarization of the $n=0$ states, which correlates with the emission of a nonzero net momentum in the direction of the magnetic field. In contrast, in higher Landau levels, states with opposite spins drive neutrino emissions with opposing net momenta. Such contributions largely, but not entirely, cancel each other out, resulting in a less pronounced net effect.

Using the illustration of the Fermi surface in figure~\ref{fig:Fermi-surface}, we see that the Fermi surface states associated with the LLL ($n=0$) are located near the polar regions, where the electron longitudinal momenta are largest in magnitude, $|p_{e,z}|\simeq \mu_e$. As the Landau-level index $n$ increases, the Fermi surface states move closer to the equator, where $|p_{e,z}|\simeq 0$. Referring to figure~\ref{fig.rates_LLs-mue40} now, we see that contributions of Landau levels to the momentum emission decrease as $n$ increases. Moreover, the sign of partial contributions to the neutrino longitudinal momentum changes, reaching the maximum negative value when $n$ approximately equals the integer part of $\mu_e^2/(2|eB|)$. 

Generally, the positive contributions from the low-lying Landau levels outweigh the negative contributions from the higher Landau levels, leading to a net positive result for $\dot{\cal P}_{\nu,z}$, as shown previously in figure~\ref{fig.rates_20-600-T1mue40}. However, under specific conditions, such as sufficiently low temperatures and strong magnetic fields, a large negative contribution from the Landau level whose energy minimum approximately matches the Fermi-level threshold can dominate the net momentum emission. In such cases, the net result for $\dot{\cal P}_{\nu,z}$ becomes negative, as seen around $B=6.8\times 10^{16}~\mbox{G}$ and $B=1.35\times 10^{17}~\mbox{G}$ in figure~\ref{fig.rates_20-600-T1mue40} for the lowest temperature case, $T=0.5~\mbox{MeV}$.

\subsection{Comparison of rates at different electron chemical potentials}

\begin{figure}[t]
\centering
  \subfigure[]{\includegraphics[width=0.45\textwidth]{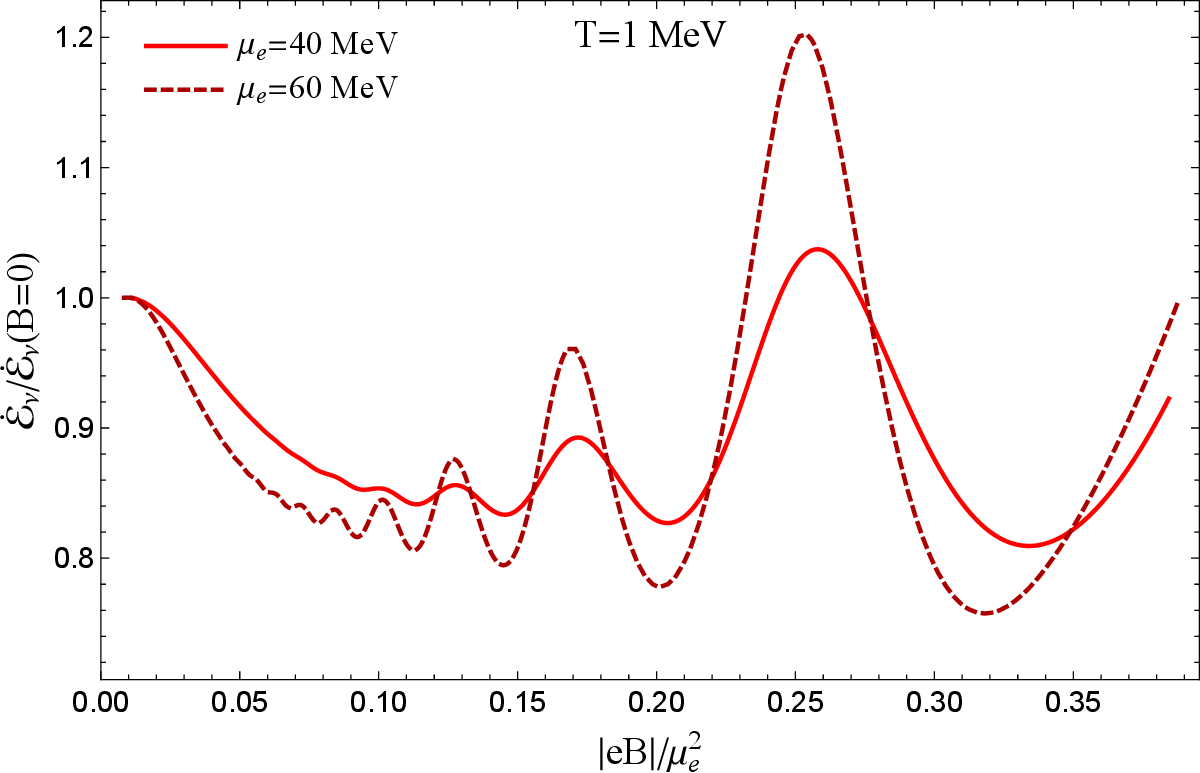}}
  \hspace{0.05\textwidth}
  \subfigure[]{\includegraphics[width=0.45\textwidth]{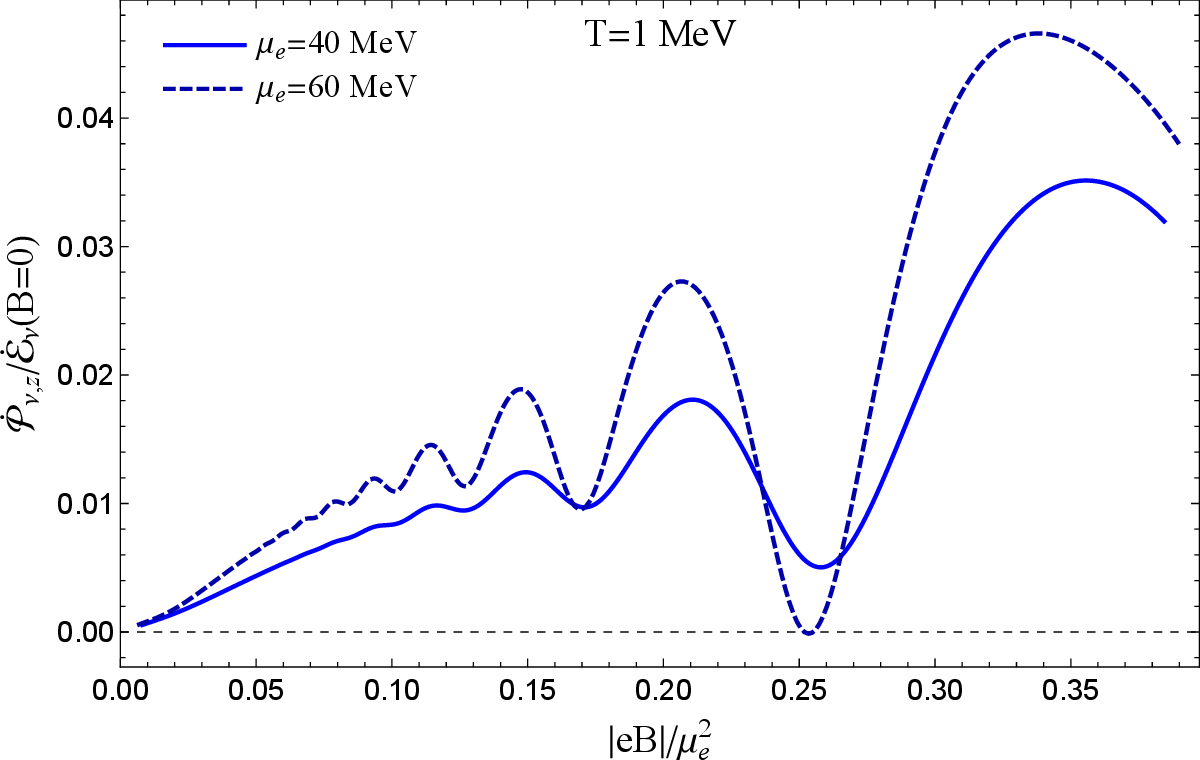}}
\caption{The neutrino energy (a) and momentum (b) emission rates for two different choices of the 
electron chemical potential: $\mu_e=40~\mbox{MeV}$ (solid lines) and $\mu_e=60~\mbox{MeV}$ (dashed lines).}
\label{fig.rates-mue40-60}
\end{figure}

To better understand how the rates depend on the model parameters, here we examine how the results change with variations in the electron chemical potential. To this end, figure~\ref{fig.rates-mue40-60} compares numerical data for the same temperature but two different values of the chemical potential: $\mu_e = 40~\mbox{MeV}$ (solid lines) and $\mu_e = 60~\mbox{MeV}$ (dashed lines). The rates are plotted as functions of the dimensionless ratio $|eB|/\mu_e^2$, which offers the advantage of aligning the points where the Fermi surface matches the Landau-level thresholds.

As seen in figure~\ref{fig.rates-mue40-60}, the magnitude of oscillations is greater for  $\mu_e = 60~\mbox{MeV}$. This behavior is expected because, as argued earlier, the characteristic magnetic field strength for resolving oscillations is governed by the condition $|eB|\gtrsim \pi T \mu_e$. When expressed in terms of the dimensionless ratio on the horizontal axis, this condition becomes $|eB|/\mu_e^2\gtrsim \pi T /\mu_e$, which decreases as $\mu_e$ increases. Therefore, a larger $\mu_e$ implies that the region of well-resolved oscillations is extended to lower values of $|eB|/\mu_e^2$. Additionally, the magnitude of oscillations for $\mu_e = 60~\mbox{MeV}$ is greater than for $\mu_e = 40~\mbox{MeV}$. 

Note that in the right panel of figure~\ref{fig.rates-mue40-60}, the momentum emission rate is again normalized by the energy emission rate at $B=0$. Alternatively, it can also be normalized by the energy emission rate at a nonzero $B$. While the specific choice of normalization affects the quantitative results, the impact is relatively minor. This is because the rate's dependence on the magnetic field is quite weak. As shown in the left panel of figure~\ref{fig.rates-mue40-60}, the average rate is suppressed by only about $10\%$ to $20\%$ overall when the oscillations are disregarded.

By comparing our numerical data for the energy and momentum emission rates, we estimate their ratio to be approximately
\begin{eqnarray}
\eta \equiv \frac{ \dot{\cal P}_{\nu,z}}{\dot{\cal E}_{\nu}} \sim 2 \times 10^{-3} \frac{|eB|}{\mu_e T}.
\label{def-eta}
\end{eqnarray} 
This implies that the asymmetry in momentum emission reaches a level of about $1.5\%$ only when the magnetic field exceeds approximately $5\times 10^{16}~\mbox{G}$, assuming $\mu_e= 40~\mbox{MeV}$ and $T= 1~\mbox{MeV}$. Even at magnetic fields so strong that all occupied electron states are confined to the LLL, the peak value remains moderate, $\eta\lesssim 0.15$. In comparison, this is smaller than the values obtained in simple models relying on the electron spin polarization to roughly estimate neutrino emission asymmetry \cite{Sagert:2007as,Ayala:2018kie,Ayala:2024wgb}.

\subsection{Interplay of Fermi-liquid corrections and magnetic fields}

It is important to reiterate that the Fermi-liquid effects in dense quark matter are crucial for accurately deriving the Urca rates \cite{Iwamoto:1980eb,Iwamoto:1982zz}. Without such corrections, the phase space for the relevant weak processes becomes severely restricted, requiring the electron and both quark momenta to be nearly parallel. This condition, in turn, suppresses the rate of the direct Urca processes by a factor of $T/\mu_f$ \cite{Burrows:1980ec}.

One might expect that the background magnetic field naturally resolves the phase space restriction even without the inclusion of the Fermi-liquid corrections. Indeed, it effectively relaxes momentum conservation condition because the transverse components of the electron momentum are not conserved quantum numbers any longer. 

\begin{figure}[t]
\centering
  \subfigure[]{\includegraphics[width=0.45\textwidth]{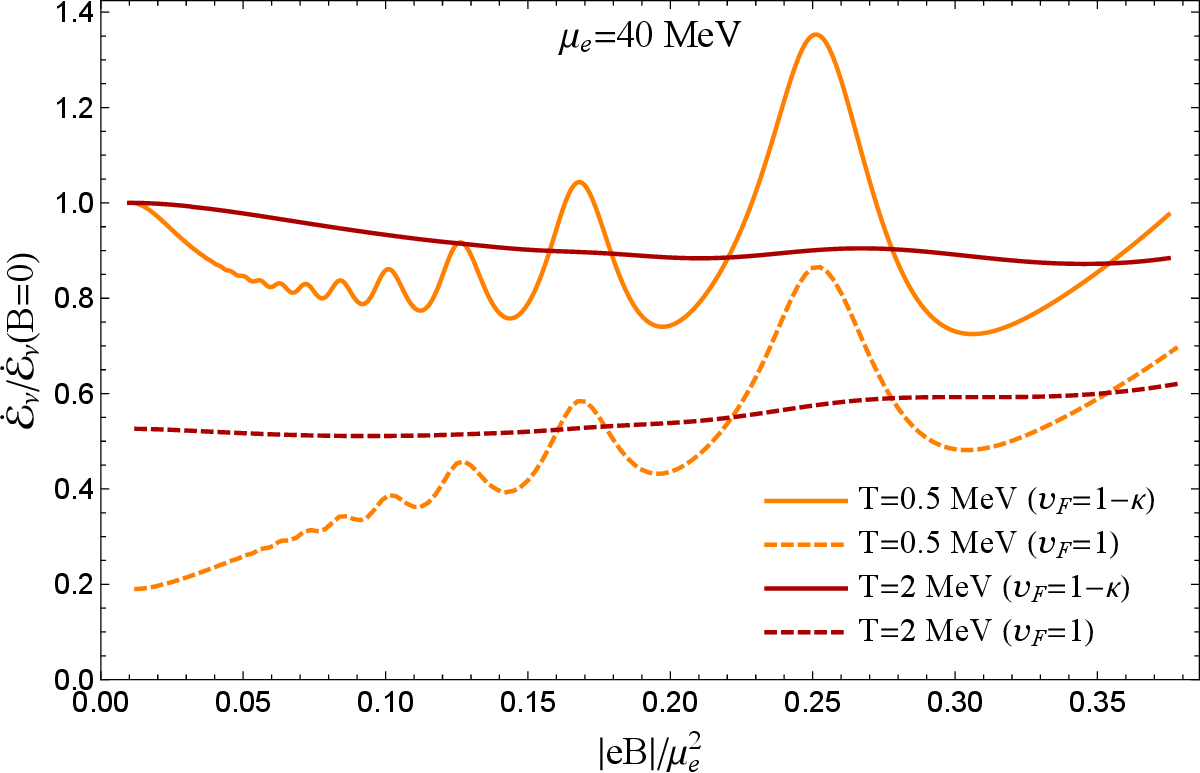}}
  \hspace{0.05\textwidth}
  \subfigure[]{\includegraphics[width=0.45\textwidth]{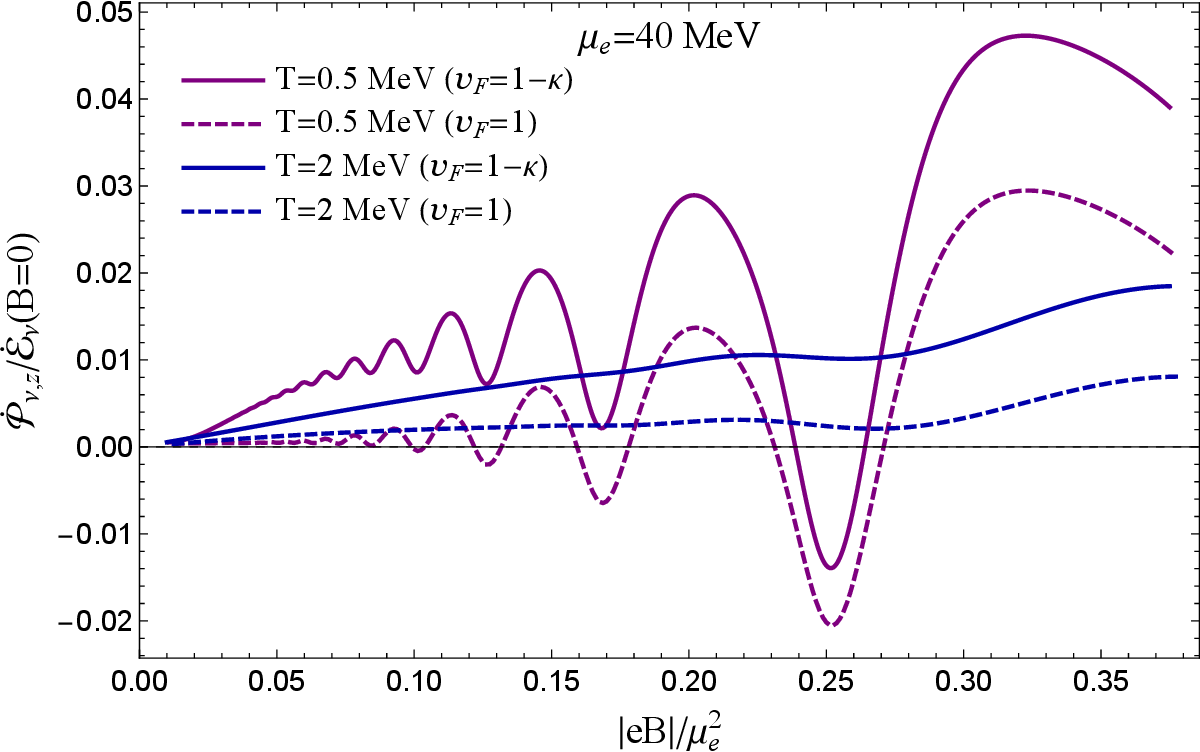}}
\caption{The neutrino energy (panel a) and momentum (panel b) emission rates with Fermi-liquid corrections (solid lines) and without Fermi-liquid corrections  (dashed lines) for two different temperatures: $T=0.5~\mbox{MeV}$ and $T=2~\mbox{MeV}$.}
\label{fig.rates-noFL}
\end{figure}

To test this premise, we compare the results for the rate with and without the Fermi-liquid corrections in figure~\ref{fig.rates-noFL}, where we present two sets of data for $T=0.5~\mbox{MeV}$ and $T=2~\mbox{MeV}$, normalized by the zero-field rate in Eq.~(\ref{dot-E-CT-app-B0}). The rates with Fermi-liquid effects are represented by solid lines, while the rates without such effects are shown by dashed lines. The latter are clearly systematically suppressed in comparison to the former.

For weak magnetic fields, we see that the energy rates are much smaller when the Fermi-liquid corrections are omitted. The suppression factor is roughly on the order of $T/(\kappa \mu_f)$, which is qualitatively consistent with the result in ref.~\cite{Burrows:1980ec}. Note that an extra factor of $\kappa$ appears in the suppression because Iwamoto's result is itself proportional to $\kappa $. The difference between the numerical results with and without the Fermi-liquid effects, shown in figure~\ref{fig.rates-noFL}(a), becomes increasingly negligible for very strong magnetic fields. This observation supports the hypothesis that the magnetic field reduces the sensitivity to Fermi-liquid corrections, as it serves a similar role.

The behavior of the net longitudinal momentum emission is more subtle in figure~\ref{fig.rates-noFL}(b). When the quark Fermi-liquid effects are excluded, the average momentum emission rate remains significantly suppressed, even in the presence of relatively strong fields. This outcome is perhaps not surprising, given that the relaxation of particle momentum collinearity by the magnetic field primarily affects the transverse motion of electrons. The corresponding states are located near the equatorial region of the Fermi surface (see Fig.\ref{fig:Fermi-surface}) and predominantly contribute to a negative value of $\dot{\cal P}_{\nu,z}$. Conversely, positive contributions to $\dot{\cal P}_{\nu,z}$ arise mainly from the polar regions of the Fermi surface (see Fig.\ref{fig:Fermi-surface}), where Fermi-liquid effects likely remain critical for significantly enhancing the emission rate. This contrast between the equatorial and polar contributions explains the numerical results in figure~\ref{fig.rates-noFL}(b), where the momentum emission rate is substantially reduced on average when Fermi-liquid effects are absent.

\subsection{Estimates of pulsar kick velocity}

Using the neutrino emission rates obtained in the preceding subsections, here we explore their potential implications for strongly magnetized compact stars.

The average energy emission rate decreases with increasing magnetic field strength, indicating that strong magnetic fields might slightly slow stellar cooling. However, the suppression remains modest, with a maximum reduction of about 20\% even at extreme field strengths of $B\sim 10^{17}~\mbox{G}$, making significant observational effects unlikely. An exception could arise in the LLL regime, where the energy emission rate would increase substantially, potentially accelerating cooling. However, this regime appears improbable under typical stellar conditions.

A nonzero net momentum emission along the magnetic field direction raises the possibility of neutrino emissions contributing to pulsar kicks. The corresponding estimate for the kick velocity $v_k$ is given by
\begin{eqnarray}
 v_k=\frac{4\pi R_c^3}{3 M}\int \dot{\cal P}_{\nu,z}dt
 =\frac{4\pi R_c^3}{3 M} \int \eta \,  \dot{\cal E}_{\nu}dt ,
 \label{v-kick-def}
 \end{eqnarray} 
where $R_c$ is the radius of the quark core and $M$ is the mass of a neutron star. In the last form of Eq.~(\ref{v-kick-def}), we expressed the momentum emission rates in terms of the corresponding energy rate by using function $\eta $ that was introduced in Eq.~(\ref{def-eta}). 

Assuming that the neutrino emission is the main cooling mechanism, the corresponding rate $\dot{\cal E}_\nu$ determines the cooling rate of the stellar quark core, 
\begin{eqnarray}
\frac{dT}{dt}=-\frac{\dot{\cal E}_\nu}{C_v}.
\label{cooling-rate}
\end{eqnarray}
where $C_v$ is the specific heat of dense quark matter. Its analytical expression is given by~\cite{Glendenning:1997wn}
\begin{eqnarray}
C_v(T)= 3 (\mu_u^2+\mu_d^2) T\left(1-\frac{2\alpha_s}{\pi}\right) +O\left(T^3\right),
\end{eqnarray}
with subleading terms suppressed by a factor on the order of $T^2/\mu_f^2$. Note that the electron contribution to the specific heat is neglected, as it is suppressed by a factor of approximately $(\mu_e/\mu_u)^2$. Using the relation in Eq.~(\ref{cooling-rate}), we can change the integration variable in Eq.~(\ref{v-kick-def}) from time $t$ to temperature $T$, allowing us to derive 
\begin{eqnarray}
 v_k  = - \frac{4\pi R_c^3}{3 M} \int_{T_i}^{T_f} \eta \, C_v(T')dT^\prime .
\end{eqnarray} 
Finally, using the estimate for function $\eta $ in Eq.~(\ref{def-eta}), we obtain
\begin{eqnarray}
 v_k\simeq 8\pi  \times 10^{-3} \frac{R_c^3}{M_{\odot}} \frac{|eB|}{\mu_e} (\mu_u^2+\mu_d^2) \left(1-\frac{2\alpha_s}{\pi}\right)(T_i-T_f).
\label{v-kick-eq}
\end{eqnarray} 
The corresponding numerical estimate reads
\begin{eqnarray}
 v_k \simeq 1.9~\mbox{km/s}\, \left(\frac{B}{10^{16}~\mbox{G}}\right)
 \left(\frac{M_{\odot}}{M}\right) \left(\frac{R_c}{10~\mbox{km}}\right)^3
 \left(\frac{40~\mbox{MeV}}{\mu_e}\right)
 \left(\frac{\mu_f}{300~\mbox{MeV}}\right)^2\left(\frac{\Delta T}{10~\mbox{MeV}}\right),
  \label{v-kick-num}
\end{eqnarray} 
where $M_{\odot}$ is the solar mass. For simplicity, here we used a common quark chemical potential $\mu_f$.

We note in passing that because of high Urca emission rate, the dense quark matter should cool quite fast. This is confirmed by the following estimate for the cooling time:
\begin{eqnarray}
 \Delta t =-\int_{T_i}^{T_f} \frac{C_v(T')}{\dot{\cal E}_\nu(T')}\,dT'
 \simeq \frac{945}{457\alpha_s G_F^2\cos^2\theta_C\mu_e T_f^4} \left(1-\frac{2\alpha_s}{\pi}\right),
\end{eqnarray}
where for simplicity we used Iwamoto's rate (\ref{dot-E-app-Iwamoto}), replaced both $\mu_u$ and $\mu_d$ with a common quark chemical potential $\mu_f$, and ignored the weak dependence on the initial temperature. Numerically, this gives
\begin{eqnarray}
 \Delta t \simeq 119~\mbox{min}\, \left(\frac{40~\mbox{MeV}}{\mu_e}\right) \left(\frac{0.1~\mbox{MeV}}{T_f}\right)^4 ,
\end{eqnarray}
which is indeed a relatively short time.

Our estimate for the pulsar kick velocity in Eq.~(\ref{v-kick-num}) is considerably lower than the optimistic estimates in refs.~\cite{Sagert:2007as,Ayala:2018kie,Ayala:2024wgb}, which were based on simplified models that used the electron spin polarization to estimate kick velocities. The main difference comes from  the ratio of $\dot{\cal P}_{\nu,z} $ to $\dot{\cal E}_{\nu}$, as presented in Eq.~(\ref{def-eta}). Our first-principles calculations indicate that this ratio remains quite small, even in the presence of very strong magnetic fields. This relatively small net momentum emission can be attributed to the kinematics of weak processes, where only certain regions of the Fermi surface contribute to momentum emission in the direction of the magnetic field, while other regions counterbalance it by emitting in the opposite direction.

Before concluding this section, we note the possibility of an alternative mechanism for pulsar kicks during the early deleptonization phase. In such a scenario, trapped neutrinos diffuse through strongly magnetized dense quark matter at rates that vary with direction relative to the magnetic field. Because of higher average neutrino energies, this mechanism could be quite promising. However, its detailed discussion is beyond the scope of this work and will be addressed elsewhere.

\section{Summary}
\label{sec:Summary}

In this study, we used first-principles field theory methods to calculate neutrino emission from strongly magnetized, unpaired dense quark matter, which is one of hypothetical states of matter thought to exist in the interiors of compact stars. Noting that the quark chemical potentials are much larger than those of electrons, only the Landau-level quantization of electron states was considered. We utilized the Kadanoff-Baym formalism to derive the neutrino number production rate and then used it to calculate both energy and momentum emission rates. 

The Landau-level discretization of electron states at the Fermi surface leads to an oscillatory dependence of the neutrino emission rate on the magnetic field strength, provided the field is sufficiently strong, $|eB|\gtrsim \pi T \mu _e$. These oscillations grow in magnitude as the temperature decreases and become formally divergent as $T\to 0$. The divergencies in the emission rates occur at the points where the Landau-level thresholds $\sqrt{2n|eB|}$ match exactly the Fermi energy $\mu_e$. For sufficiently weak fields or high temperatures, the thermal effects wash away oscillations in the rates. For example, assuming realistic magnetic fields below about $10^{17}~\mbox{G}$, the oscillations nearly vanish for $T\gtrsim 2~\mbox{MeV}$. 

Beyond the oscillatory behavior, the average energy emission rate decreases with increasing magnetic field strength (up to about $10^{17}~\mbox{G}$), but the suppression is modest, around $20\%$ at $T\simeq 0.5~\mbox{MeV}$, and even less at higher temperatures. In contrast, neutrino momentum emission along the magnetic field direction generally increases with field strength. The ratio of momentum to energy emission rates, $\eta \equiv \dot{\cal P}_{\nu,z}/\dot{\cal E}_{\nu}$, remains relatively small, reaching a few percent for fields near $10^{16}~\mbox{G}$ at $T\gtrsim 2~\mbox{MeV}$. Numerical estimates suggest that $\eta \simeq 2 \times 10^{-3} |eB|/(\mu_e T)$, which is relatively small compared to predictions from simplified models in refs.~\cite{Sagert:2007as,Ayala:2018kie,Ayala:2024wgb}. We argue, therefore, that neutrino momentum emission from dense quark matter is unlikely to account for even modest pulsar kicks with $v_k\simeq 100~\mbox{km/s}$.

We calculate the rates of the Urca processes $u+ e^- \rightarrow d+\nu_e$ and $d\rightarrow u +e^- + \bar{\nu}_e$ using an approach similar to Iwamoto’s \cite{Iwamoto:1980eb,Iwamoto:1982zz} by incorporating the Fermi-liquid effects of quarks to relax collinearity constraints on particle momenta. These effects are crucial in the absence of a magnetic field, as they enhance the rates significantly, by a factor proportional to $\mu_f/T$. Under strong magnetic fields, however, the non-conservation of the transverse momentum of electrons partially reduces the significance of these Fermi-liquid effects.

In conclusion, strong magnetic fields significantly modify neutrino emission rates in dense quark matter. The emission rates exhibit oscillatory behavior with field strength, and the oscillation magnitude increases as temperature decreases. Despite these modifications, the average energy emission rate is only mildly suppressed (by about $20\%$) even for very strong fields, suggesting limited impact on the energy balance and cooling of quark stars. While the magnetic field introduces asymmetry in neutrino emission, the resulting net momentum emission rate appears insufficient to produce substantial kicks in compact stars.

In future research, we would like to investigate whether the magnetic field's influence on momentum emission can be more pronounced in other phases of quark matter. Indeed, certain phases, particularly those involving diquark pairing and additional broken symmetries, may exhibit fundamentally different neutrino emission properties, potentially leading to significant changes in both the rates and spatial anisotropies of emission. These effects could, in turn, enhance the net momentum carried away by neutrinos, resulting in substantially larger momentum kicks. However, without a detailed exploration of all realistic quark matter phases, it remains unclear whether magnetic field-driven pulsar kicks are feasible.

Since this study focused on dense quark matter without considering neutrino trapping, it did not address the mechanism of anisotropic neutrino diffusion in strongly magnetized dense quark matter during the early deleptonization phase. In the future, it would be valuable to investigate this regime in detail, providing estimates for the resulting momentum asymmetry, its dependence on magnetic field strength, and its potential contribution to the generation of pulsar kicks.

\acknowledgments

This research was funded in part by the U.S. National Science Foundation under Grant No.~PHY-2209470.

\appendix

\section{Matsubara summation}
 \label{app:Matsubara}
 
In this Appendix, we derive a general master formula for the Matsubara summation, which is used in calculating the retarded self-energy of the W-boson at the leading one-loop order, i.e.,
\begin{eqnarray}
S(X,Y,Z,W)&=&T\sum_{k=-\infty}^{\infty} 
\frac{X (i\omega_k+\mu_a)(i\omega_k-i\Omega_m+ \mu_b)
+Y(i\omega_k+\mu_a)
+Z(i\omega_k-i\Omega_m+ \mu_b)
+W}{\left[(\omega_k-i \mu_a)^2+a^2\right]\left[(\omega_k-\Omega_m-i\mu_b)^2+b^2\right]} 
\nonumber\\
&=& \sum_{\lambda,\eta=\pm}  \frac{abX+a \eta \lambda Y+b \eta Z+\lambda W}{4ab}
\frac{n_F(a-\eta\mu_a)-n_F(\lambda b-\eta\mu_b)}{a-\lambda b-\eta (i \Omega_m +\mu_a-\mu_b)},
\label{Matsubara-sum-master}
\end{eqnarray}
where $n_F(E)=1/\left[\exp(E/T)+1\right]$ is the Fermi-Dirac distribution function. By assumption, functions $X$, $Y$, $Z$, and $W$ are independent of the Matsubara frequencies $\omega_k =(2k+1)\pi T$ and $\Omega_m=2m\pi T$. In the context of the W-boson self-energy, $a$ and $b$ are the energies of the up and down quarks, while $\mu_a$ and $\mu_b$ are their chemical potentials.

\section{Useful table integrals}
 \label{app:integrals}

In this Appendix, we present several table integrals used in deriving the neutrino emission rates in the main text, i.e.,
\begin{eqnarray}
I_1&=&\int _{-\infty}^\infty \frac{dx}{\left(e^{a-x} +1\right)\left(e^{x} +1\right)}=\frac{a}{e^{a} -1} ,
\label{int_01app} \\
I_2&=&\int _{-\infty}^\infty \frac{x dx}{\left(e^{a-x} +1\right)\left(e^{x} -1\right)}  = \frac{a^2+\pi^2}{2(e^{a} +1)} ,
\label{int_02app}\\
I_3&=&\int_{0}^{\infty} x^3 d x \frac{x^2+\pi^2}{e^{x}+1} =  \frac{457}{2520} \pi^6 , 
\label{int_03app} \\
I_4&=&\int_{-\infty}^{\infty} \frac{d u}{ e^{u}+1 }
\left( \mbox{Li}_{5}\left(e^{u} \right) -\frac{u}{4} \mbox{Li}_{4}\left(e^{u} \right)\right)
=\frac{457}{120960} \pi^6  
\label{int_04app}, \\
I_5&=&\int_{0}^{\infty} x^3 dx  \frac{(x-a)}{e^{x-a}-1}
=24 \left( \mbox{Li}_{5}\left(e^{a} \right) -\frac{a}{4}\mbox{Li}_{4}\left(e^{a} \right)\right) \nonumber\\
&=& 24 \left( \mbox{Li}_{5}\left(e^{-a} \right) +\frac{a}{4}\mbox{Li}_{4}\left(e^{-a} \right)\right) +\frac{a^5}{20}+\frac{\pi^2 a^3}{3}+\frac{2\pi^4 a}{5}.
\label{int_05app}
\end{eqnarray}
Note that the last integral has two equivalent forms, producing real results for all real values of $a$. However, since the polylogarithm function $\mbox{Li}_{n}(z)$ becomes complex when its argument exceeds $1$, the first form is more suitable for $a<0$, while the second is more convenient for $a>0$.
 
Using the above table integrals, one can also obtain the following triple integral:
\begin{equation}
I_6 =  \int_{0}^{\infty} w^3  dw \int_{-\infty}^{\infty} \int_{-\infty}^{\infty} \frac{du  d v}
{\left(e^{u}+1\right)\left(e^{v}+1\right)\left(e^{w-u-v}+1\right)}
=\frac{457}{5040} \pi^6 .
\label{int_06app}
\end{equation}

\section{Electron propagator and its spectral density}
\label{app:E-propagator}

In the presence of a nonzero magnetic field and a nonzero chemical potential ($\mu_e$), the electron propagator has the following structure in coordinate space \cite{Schwinger:1951nm}:
\begin{equation}
S_e (u,u^{\prime})=e^{i\Phi(u,u^{\prime})}\bar{S}_e (u-u^{\prime}),
\end{equation}
where $\Phi(u,u^{\prime})$ represents the well-known Schwinger phase and $\bar{S}_e (u-u^{\prime})$ denotes the translationally invariant part of the propagator. For calculations,  it is often more convenient to work with the Fourier transform of $\bar{S}_e (u-u^{\prime})$, which has the following explicit form in the Landau-level representation \cite{Miransky:2015ava}:
\begin{eqnarray}
\bar{S}_e (p_0,\bm{p})&=& i e^{-p_\perp^2\ell^2} \sum_{n=0}^{\infty} \sum_{\lambda=\pm} \frac{(-1)^n}{E_{e,n}
\left[p_0+\mu_e+i\epsilon \, \sign(p_0)-\lambda E_{e,n}\right]}
\Big\{\left[E_{e,n} \gamma^0+\lambda(m_{e}-p_{e,z}\gamma^3)\right]\nonumber\\
&& \times \left[{\cal P}_{+}L_n\left(2 p_\perp^2\ell^2\right)
-{\cal P}_{-}L_{n-1}\left(2 p_\perp^2\ell^2\right)\right]
+2\lambda (\bm{p}_\perp\cdot\bm{\gamma}_\perp) L_{n-1}^1\left(2 p_\perp^2 \ell^2\right)\Big\} .
  \label{prop-elelctron}
\end{eqnarray}
Here $ E_{e,n}=\sqrt{2n|e B|+p_{e,z}^2+ m_e^2 }$ are the Landau-level energies, $\ell =1/\sqrt{|eB|}$ is the magnetic length for the electron, ${\cal P}_{\pm}=(1\pm i s_\perp \gamma^1\gamma^2)/2$ are the spin projectors, $s_\perp = \mbox{sign}(e B)$, and $L_n^{\alpha}\left(z\right)$ are the generalized Laguerre polynomials.

In this study, we adopt the following convention for the spectral representation of the fermion propagators: 
\begin{equation}
\bar{S}_{e}(p_0,\bm{p})  =  i  \int_{-\infty}^{\infty} \frac{dk_{0}}{2\pi} \frac{ A_{e} (k_0+\mu_{e},\bm{p}) }{p_0-k_{0}},
  \label{prop-spectral-fun}
 \end{equation}
 where the spectral function $A_{e} (p_0,\bm{p})$ 
is defined as the difference between the retarded and advanced propagators at $\mu_e=0$, i.e.,
 \begin{equation}
A_{e} (p_0,\bm{p})= \left[ \bar{S}_{e}(p_0+i0,\bm{p}) - \bar{S}_{e}(p_0-i0,\bm{p})\right]\Big|_{\mu_{e}=0}.
\label{spectral-density-app0}
\end{equation}
In the case of the electron propagator in Eq.~\eqref{prop-elelctron}, the spectral function is
\begin{eqnarray}
A_{e} (p_0,\bm{p})&=& 2\pi e^{-p_\perp^2\ell^2} \sum_{n=0}^{\infty} \sum_{\lambda=\pm} \frac{(-1)^n}{E_{e,n} }
\Big\{\left[E_{e,n} \gamma^0+\lambda(m_{e}-p_{e,z}\gamma^3)\right]\left[{\cal P}_{+}L_n\left(2 p_\perp^2\ell^2\right)
-{\cal P}_{-}L_{n-1}\left(2 p_\perp^2\ell^2\right)\right]
\nonumber\\
&&+2\lambda (\bm{p}_\perp\cdot\bm{\gamma}_\perp) L_{n-1}^1\left(2 p_\perp^2 \ell^2\right)\Big\}
\delta\left( p_0-\lambda E_{e,n}\right).
\label{spectral-density-app1}
\end{eqnarray}
For comparison, the zero-field expressions for the electron propagator and the spectral function read
\begin{eqnarray}
S_e^{(0)}(p_0,\bm{p}) &=& i \frac{(p_0+\mu_e )\gamma^0- (\bm{p}\cdot\bm{\gamma})+m_e }{[p_0+\mu_e +i\epsilon \, \sign(p_0)]^2 -E_{e,p}^2} ,
\\
A_e^{(0)} (p_0,\bm{p}) &=&  \frac{\pi}{E_{e,p}} \sum_{\lambda=\pm} \left[E_{e,p} \gamma^0- \lambda (\bm{p}\cdot\bm{\gamma})+\lambda m_e  \right]\delta \left(p_0- \lambda E_{e,p}\right),
\end{eqnarray}
where $E_{e,p}= \sqrt{p^2+m_e^2}$.

\section{Lorentz contraction of the lepton and quark tensors}
\label{sec-L-Pi}

In this appendix, we derive an explicit expression for the contraction of the lepton and quark tensors that appears in the neutrino rate in Eq.~\eqref{rate-01}, i.e., 
\begin{equation}
\mbox{Im}\left[F(Q,P_\nu)\right] = 
L_{n,\lambda}^{\delta\sigma}(\bm{p}_e,\bm{p}_\nu)  \Im \left[ \Pi^R_{\delta\sigma}(Q) \right] ,
\end{equation}
where $Q\equiv(E_{e,n}-p_\nu -\mu_e,\bm{p}_e-\bm{p}_\nu)$. 

The lepton tensor, which captures the Dirac (spin) structure of the electron and neutrino, is given by 
\begin{eqnarray}
L^{\delta\sigma}_{n,\lambda}(\bm{p}_e,\bm{p}_\nu)&=&\mbox{Tr}\Big[\Big\{
\left(E_{e,n}\gamma^{0} -\lambda  p_{e,z}\gamma^3+ \lambda m_e \right) 
\left[{\cal P}_{+}L_n\left(2 p_{e,\perp}^2\ell^2\right)
-{\cal P}_{-}L_{n-1}\left(2 p_{e,\perp}^2\ell^2\right)\right] \nonumber\\
&&
+2\lambda  (\bm{p}_{e,\perp}\cdot\bm{\gamma}_\perp) L_{n-1}^1\left(2 p_{e,\perp}^2\ell^2\right)
\Big\}\gamma^\sigma(1-\gamma^5)(\gamma_0 p_\nu-\bm{\gamma}\cdot \bm{p}_\nu)\gamma^\delta (1-\gamma^5)\Big].
\label{Lmunu-app}
\end{eqnarray}
In the case of dense quark matter, the retarded self-energy of the gauge boson is determined by the simplest one-loop quark diagram, as illustrated in figure~\ref{fig.NuSelfEnergy}b. The corresponding expression reads
\begin{equation}
\Pi_{R}^{\delta\sigma}(q_0,\bm{q}) = - i N_c  \sumint \frac{d^4K}{(2\pi)^4}\tr\left[\gamma^\delta(1-\gamma_5) \bar{S}_u(K)\gamma^\sigma(1-\gamma_5) \bar{S}_d(P)\right],
\label{pimunu-app}
\end{equation}
where $P=K+Q$ and $N_c=3$ is the number of quark colors. Nonzero temperature effects are incorporated using the imaginary-time formalism, where the integration over the energy $k_0$ is replaced with a Matsubara sum, i.e.,
\begin{equation}
\sumint \frac{d^4 K}{(2\pi)^4} f(k_0,\bm{k}) = T\sum_{k=-\infty}^{\infty}  i  \int \frac{d^3 \bm{k}}{(2\pi)^3} f(i\omega_k,\bm{k}).
\label{int-Matsubara}
\end{equation}
By definition, the fermionic Matsubara frequencies are $\omega_k=(2k+1)\pi T$. When calculating the sum, the external gauge-boson energy $q_0+i\epsilon$ is replaced with $i \Omega_m$, where $\Omega_m=2m\pi T$ is the bosonic Matsubara frequency. The dependence on $q_0$ is then restored at the end by performing the analytical continuation $i \Omega_m \to q_0+i\epsilon$.

In the presence of a background magnetic field, one should formally use the translation invariant parts $\bar{S}_f(K)$ of the quark propagators in Eq.~\eqref{pimunu-app}. However, as argued in the main text, it suffices to use an approximation that neglects the effect of the magnetic field on quarks. Therefore, here we use the zero-field quark propagators, namely
\begin{equation}
\bar{S}_f(K) = i \frac{K_\alpha \gamma^\alpha+\gamma^0 \mu_f-m_f}{(k_0+\mu_f)^2-\bm{k}^2-m_f^2},
\end{equation}
with $f=u,d$. After evaluating the Dirac trace, we obtain the following result:
\begin{equation}
  \Pi_{R}^{\delta\sigma}(q_0,\bm{q}) = 8  i  N_c \sumint \frac{d^4K}{(2\pi)^4} \frac{ (\bar{K}^\delta \bar{P}^\sigma+\bar{K}^\sigma \bar{P}^\delta)-g^{\delta\sigma}\bar{K}\cdot \bar{P}+i \epsilon^{\delta\sigma\mu\nu}\bar{K}_\mu \bar{P}_\nu }{\Delta(\bar{P})\Delta(\bar{K})},
\label{pimunu-app-d6}
\end{equation}
where $\bar{K}\equiv (k_0+\mu_f,\bm{k})$ and $\Delta(\bar{K})=(k_0+\mu_f)^2-\bm{k}^2-m_f^2$.
 
From Eqs.~\eqref{Lmunu-app} and \eqref{pimunu-app}, we derive the following result for the Lorentz contraction of the lepton and quark tensors:
\begin{equation}
F(Q,P_\nu)\equiv
L_{\delta\sigma}(\bm{p}_e,\bm{p}_\nu) \Pi_{R}^{\delta\sigma}(q_0,\bm{q})
= 128 i N_c \sumint \frac{d^4K}{(2\pi)^4}
\frac{\left(\bar{P}\cdot P_\nu\right) \left(\bar{K}\cdot Y_e\right)}{ \Delta(\bar{P}) \Delta(\bar{K})},
\label{Lpimunu-app-d7}
\end{equation}
where the explicit components of the electron four-vector $Y_e$ are given by
\begin{eqnarray}
Y_{e,0} &=& \left(E_{e,n}-s_\perp \lambda p_{e,z}\right)L_{n}\left(2 p_{e,\perp}^2\ell^2\right) 
-\left(E_{e,n}+s_\perp \lambda p_{e,z}\right)L_{n-1}\left(2 p_{e,\perp}^2\ell^2\right) ,
\label{Y0-app}\\
Y_{e,z}&=&\left(\lambda p_{e,z}-s_\perp E_{e,n}\right)L_{n}\left(2 p_{e,\perp}^2\ell^2\right) 
-\left(\lambda p_{e,z}+s_\perp  E_{e,n}\right)L_{n-1}\left(2 p_{e,\perp}^2\ell^2\right) 
\label{Yz-app},\\
\bm{Y}_{e,\perp} &=& -4\lambda \bm{p}_{e,\perp} L_{n-1}^1\left(2 p_{e,\perp}^2\ell^2\right).
\label{Yxy-app}
\end{eqnarray}
In passing, we note that the zero-field expression analogous to Eq.~(\ref{Lpimunu-app-d7}) contains a similar combination of the particle four-momenta, $(\bar{P}\cdot P_\nu)(\bar{K}\cdot P_e)$, in the numerator of its integrand. Up to an overall constant, this combination coincides with the squared scattering amplitude, $|{\cal M}|^2$, of the direct Urca processes. 

In the presence of a nonzero field, as seen from Eq.~(\ref{Lpimunu-app-d7}), the momentum-dependent function that defines the squared scattering amplitude is replaced by:
\begin{eqnarray}
\left(\bar{P}\cdot P_\nu\right) \left(\bar{K}\cdot Y_e\right)
& = & \left[ (p_0+\mu_d) p_{\nu,0}  - \bm{p}\cdot \bm{p}_{\nu}\right]\big[
(k_0+\mu_u+s_\perp k_z)\left(E_{e,n}-s_\perp \lambda p_{e,z}\right)L_{n}\left(2 p_{e,\perp}^2\ell^2\right)\nonumber\\
&&-(k_0+\mu_u-s_\perp k_z)\left(E_{e,n}+s_\perp \lambda p_{e,z}\right)L_{n-1}\left(2 p_{e,\perp}^2\ell^2\right)\nonumber\\
&&
+ 4\lambda \left(\bm{k}_{\perp}\cdot\bm{p}_{e,\perp}\right)L_{n-1}^1\left(2 p_{e,\perp}^2\ell^2\right) \big] .
\label{app-PPKY}
\end{eqnarray}
Using the master formula in Eq.~\eqref{Matsubara-sum-master}, we perform the Matsubara sum in Eq.~\eqref{Lpimunu-app-d7} and arrive at the following result: 
{\small \begin{equation}
F(Q,P_\nu)= -\frac{4N_c}{\pi^3}  \sum_{\lambda^\prime ,\eta^\prime=\pm} \int \frac{ d^3\bm{k}}{E_k E_p}
\frac{n_F(E_p-\eta^\prime\mu_d)-n_F(\lambda^\prime E_k-\eta^\prime\mu_u)}{E_p-\lambda^\prime E_k-\eta^\prime (q_0 +\mu_d-\mu_u+i\epsilon)} 
 (E_{p}  p_{\nu,0}-\eta^\prime \bm{p}\cdot \bm{p}_\nu)(E_{k} Y_{e,0} - \eta^\prime \lambda^\prime \bm{k}\cdot\bm{Y}_{e}) .
\label{Lpimunu-app-d12}
\end{equation}}
Noting that the quark chemical potentials are large compared to the temperature ($\mu_f \gg T $ for $f=u,d$), it is justified to keep only contributions from the quark states near the Fermi surface ($\lambda^\prime=\eta^\prime=1$) and neglect all contributions from the antiquarks. Then, the approximate result reduces down to
\begin{equation}
F(Q,P_\nu)
\simeq -\frac{4N_c}{\pi^3}  \int \frac{ d^3\bm{k}}{E_k E_p} \frac{n_F( E_p- \mu_d)-n_F(E_k-\mu_u)}{(E_p-\mu_d)- (E_k-\mu_u)- q_0-i\epsilon}
 (E_p  p_{\nu,0}-\bm{p}\cdot \bm{p}_\nu)(E_k Y_{e,0} - \bm{k}\cdot\bm{Y}_{e}) .
\end{equation}
Similarly, by neglecting the contributions of positrons in the lepton tensor, we keep only the $\lambda=1$ contribution in the electron four-vector $Y_e$, see Eqs.~\eqref{Y0-app} -- \eqref{Yxy-app}.

After performing the analytical continuation $i \Omega_m \to q_0+i\epsilon$ and using the Sokhotski formula, we extract the imaginary (absorptive) part of function $ F(Q,P_\nu) $,
\begin{eqnarray}
\mbox{Im}\left[F(Q,P_\nu)\right] &=& -\frac{4N_c}{\pi^2}  \int \frac{d^3\bm{k}}{E_k E_p}  
\left[n_F( E_p- \mu_d)-n_F(E_k-\mu_u)\right]  (E_p  p_{\nu,0}-\bm{p}\cdot \bm{p}_\nu)(E_k Y_{e,0} - \bm{k}\cdot\bm{Y}_{e}) \nonumber \\
&\times&  
 \delta\left[(E_p-\mu_d)- (E_k-\mu_u)- q_0\right] .
 \end{eqnarray}
Finally, after integrating over the angular coordinates, we obtain
{\small \begin{eqnarray}
\mbox{Im}\left[F(Q,P_\nu)\right] &=& -\frac{8N_c  }{\pi}\int \frac{pkdk}{v_Fq E_k E_p} \left[n_F( E_p- \mu_d)-n_F(E_k-\mu_u)\right] \Bigg\{ \left[ E_p  p_{\nu,0}- \left(1+ \frac{k}{q} \cos\theta_{eu}\right) (\bm{q}\cdot \bm{p}_\nu) \right]\nonumber \\
&&  \times\left( E_k Y_{e,0}  -\frac{k}{q}  \cos\theta_{eu} (\bm{q}\cdot \bm{Y}_{e}) \right)
 +\frac{1-\cos^2\theta_{eu}}{2}\left( k^2 (\bm{p}_\nu\cdot \bm{Y}_{e}) -\frac{k^2}{q^2}(\bm{p}_\nu\cdot \bm{q})(\bm{Y}_{e}\cdot \bm{q})\right) 
\Bigg\} ,
\label{L-Im-Pi}
\end{eqnarray}}
where, to perform the integration over the polar angle $\theta$, we used the following relation for the $\delta$-function
\begin{equation}
\delta\left[v_F(p-p_F)-v_F(k-k_F)- q_0\right] = \frac{v_F k q}{p} \delta\left(\cos\theta-\cos\theta_{eu}\right). 
\end{equation}
Here the analog of the zero-field results in Eq.~\eqref{costheta0} reads
\begin{equation}
\cos\theta_{eu}  \simeq \frac{p_F^2-k_F^2-q^2}{2k_Fq} \simeq \frac{v_F^2 (\mu_d^2-\mu_u^2)-p_e^2}{2v_F\mu_u p_e} .
\label{cos-theta-app}
\end{equation}
Note that the condition $|\cos\theta_{eu}| \leq 1$ imposes the constraint  $v_F\mu_e \leq p_e \leq v_F(\mu_d+\mu_u)$.
As in the zero-field case, we have accounted for the Fermi-liquid corrections in the quark dispersion relations, as described in Eqs.~\eqref{Ep-FL}--\eqref{Ek-FL}. We also assumed that the quark momenta are approximately equal to their Fermi-surface values, $p\approx p_F = v_F \mu_d$ and $k\approx k_F= v_F \mu_u$, while the neutrino momentum (with typical values of the order of temperature) is negligible compared to those of electrons and quarks. However, unlike the zero-field case, we cannot replace $p_e \equiv \sqrt{p_{e,\perp}^2+p_{e,z}^2}$ with the electron's Fermi momentum because $p_{e,\perp}^2$ is not the physical transverse momentum in the presence of a magnetic field. Recall that the Fermi surface for electrons in a given $n$-th Landau level is determined by the solutions to the equation $\sqrt{2n|eB|+p_{z,F}^2} = \mu_e$.

In  passing, let us note the following expression:
\begin{equation}
1-\cos^2\theta_{eu} \simeq \frac{\left(p_e^2-v_F^2\mu_e^2\right) \left[v_F^2(\mu_d+\mu_u)^2-p_e^2\right]}{4v_F^2 \mu_u^2 p_e^2} ,
\end{equation}
which appears as a multiplier for the second term within the curly brackets in Eq.~\eqref{L-Im-Pi}.

\section{Neutrino emission rate in the absence of a magnetic field}
 \label{app:emissionB0}
 
In this Appendix, we rederive the well-known zero-field result originally obtained by Iwamoto in the 1980s \cite{Iwamoto:1980eb,Iwamoto:1982zz}, which serves as a useful benchmark for calculating the neutron emission rate in the presence of a nonzero magnetic field. Some refinements of Iwamoto's original result, as well as its extensions to color-superconducting quark matter, are discussed in refs.~\cite{Schafer:2004jp,Jaikumar:2005hy,Schmitt:2005wg}.

In the absence of a magnetic field, the neutrino-number production rate  is given by
\begin{eqnarray}
\frac{\partial f_\nu(t,\bm{p}_\nu)}{\partial t} &=& \frac{N_c  G_F^2\cos^2\theta_C}{ 2  \pi^4}v_F \mu_u \mu_d
 \int \frac{dk d^3\bm{p}_{e} }{p_e}
n_F(E_{e}-\mu_e) 
\nonumber\\
&&\times n_F(p_{\nu}-E_k-E_{e}+\mu_d)n_F(E_k-\mu_u) \left( 1 -v_F \frac{p_e }{\mu_{e}}  \cos\theta^{(0)}_{eu}  \right) ,
\label{rate-app1-B0}
\end{eqnarray}
which is analogous to Eq.~(\ref{rate-02}) at nonzero magnetic field. Here, we account for the Fermi-liquid corrections in the quark dispersion relations, which take the following approximate forms near the Fermi surfaces \cite{Baym:1975va,Schafer:2004jp}:
 \begin{eqnarray}
E_p &\simeq& \mu_d+v_F(p-p_F), \label{Ep-FL} \\
E_k &\simeq& \mu_u+v_F(k-k_F), \label{Ek-FL} 
\end{eqnarray}
where $p_F = v_F \mu_d$, $k_F = v_F \mu_u$, and $v_F=1-\kappa$ with $\kappa= 2\alpha_s/(3\pi)$. Note that the Fermi-liquid corrections for the electron are negligible (i.e., $p_{e,F}\simeq \mu_e$) because, unlike the QCD coupling $\alpha_s$, the fine structure constant is very small.

By definition, $\theta^{(0)}_{eu}$ in Eq.~\eqref{rate-app1-B0} is the angle between the electron and up-quark momenta. Its value is determined by the following relation:
\begin{eqnarray}
\cos\theta^{(0)}_{eu}  &\simeq &\frac{v_F^2 (\mu_d^2-\mu_u^2)-p_e^2}{2v_F\mu_u p_e} 
\simeq v_F- (1-v_F^2) \frac{\mu_e}{2v_F\mu_u} 
\label{costheta0}
\end{eqnarray}
To arrive at the last approximation, we replaced $p_e$ with the electron's Fermi momentum $p_{e,F}\simeq \mu_e$. As evident, if the quark Fermi-liquid corrections were neglected (i.e., $v_F\to1$), $\cos\theta^{(0)}_{eu}$ would approach $1$. Physically, this implies that the momenta of the electron and the up quark would become collinear \cite{Iwamoto:1980eb,Iwamoto:1982zz}. Such a highly restricted phase space would result in a parametric suppression of the direct Urca process rates by a factor of $T/\mu_f$ \cite{Burrows:1980ec}.

By analyzing the integrand Eq.~\eqref{rate-app1-B0}, one finds that the corresponding integrals are primarily dominated by quasiparticles in a close vicinity of the Fermi surfaces, within an energy window of the order of the temperature. Therefore, following the original studies in ref.~\cite{Iwamoto:1980eb,Iwamoto:1982zz}, we use a conventional approximation in which the quark and electron momenta are replaced by their Fermi momenta throughout the integrand, except within the distribution functions. Then, the neutrino-number production rate reads 
\begin{eqnarray}
\frac{\partial f_\nu(t,\bm{p}_\nu)}{\partial t} &=& \frac{2N_c G_F^2\cos^2\theta_C}{ \pi^3}v_F \mu_u \mu_d \mu_e
 \int  dk dp_{e}  n_F(E_{e}-\mu_e) \nonumber\\
&&\times
n_F(p_{\nu}-E_k-E_{e}+\mu_d)n_F(E_k-\mu_u) \left( 1 -v_F \cos\theta^{(0)}_{eu} \right) .
\label{rate-app2-B0}
\end{eqnarray}
Noting that the antineutrino rate is the same, the total integrated energy rate reads
\begin{equation}
\dot{\cal E}^{(0)}_\nu (B=0) =  2 \int \frac{d^3\bm{p}_{\nu}}{(2\pi)^3} p_{\nu,0} \frac{\partial f_\nu(t,\bm{p}_\nu)}{\partial t} 
=\frac{457\pi N_c}{2520} v_F (1-v_F^2) G_F^2\cos^2\theta_C \mu_u \mu_d \mu_e T^6\left(1+\frac{\mu_e}{2\mu_u}\right) ,
 \label{dot-E-app-B0}
\end{equation}
 where we used the table integral in Eq.~\eqref{int_06app} and the definition of $\cos\theta^{(0)}_{eu} $
 in Eq.~(\ref{costheta0}). Finally, taking into account that, to leading order in $\alpha_s$, 
 \begin{equation}
v_F (1-v_F^2)  \simeq 2\kappa+O(\kappa^2)  \simeq \frac{4\alpha_s}{3\pi}+O(\alpha_s^2) ,
 \end{equation}
 we reproduce Iwamoto's well-known result, 
  \begin{equation}
 \dot{\cal E}_\nu^{\rm (Iwamoto)} \simeq\frac{457}{630} \alpha_s G_F^2\cos^2\theta_C \mu_u \mu_d \mu_e T^6 +O\left(\alpha_s^2,\frac{\mu_e}{\mu_u}\right),
  \label{dot-E-app-Iwamoto}
  \end{equation}
where we substituted $N_c=3$ and assumed that $\mu_e\ll \mu_u$. 

It should be noted that some  approximations used to derive the analytical expression for the rate in Eq.~\eqref{dot-E-app-B0} could be improved. One such improvement involves using the actual electron momentum, $p_e$, throughout the integrand in Eq.~\eqref{rate-app1-B0}, rather than approximating it with the Fermi momentum, $p_{e,F}\simeq \mu_e$. (A similar refinement for the integrals over the up and down quark momenta is unnecessary, as their corresponding Fermi momenta, $p_{d,F}$ and $p_{u,F}$, are much larger than $p_{e,F}$.) Numerical evaluation of the integral over $p_e$ with this improvement yields a slightly higher rate, which can be expressed in the following modified form:
 \begin{equation}
 \dot{\cal E}_\nu (B=0) \simeq C_T \frac{457\pi N_c}{2520}v_F (1-v_F^2)  G_F^2\cos^2\theta_C \mu_u \mu_d \mu_e T^6\left(1+\frac{\mu_e}{2\mu_u}\right),
 \label{dot-E-CT-app-B0}
\end{equation}
where $C_T$ is a function of order $1$, whose approximate dependence on the temperature is given by 
 \begin{equation}
C_T \approx 1+ c_1 \frac{T}{\mu_e}  +c_2 \frac{T^2}{\mu_e^2} , 
 \label{CT-app-B0}
\end{equation}
where $c_1\approx 15.70$ and $c_2\approx 6.287$.

\bibliographystyle{JHEP}

\providecommand{\href}[2]{#2}\begingroup\raggedright\endgroup

\end{document}